\documentclass[journal,comsoc]{IEEEtran}
\pdfoutput=1
\usepackage{epsfig}
\usepackage{subfigure}
\usepackage{calc}
\usepackage{amssymb}
\usepackage{amstext}
\usepackage{amsmath}
\usepackage{amsthm}
\usepackage{pslatex}
\usepackage{url}
\usepackage[small]{caption}
\usepackage{todonotes}

\usepackage{enumerate} 

\usepackage{lscape} 

\usepackage{algorithm}
\usepackage{algorithmicx}
\usepackage{algpseudocode}
\usepackage{pifont}

\usepackage{gensymb}

\usepackage{graphicx}
\graphicspath{ {bioimages/} }

\usepackage{tabularx,ragged2e,booktabs}

\usepackage{xcolor,colortbl}

\begin{document}


\title{RADS: Real-time Anomaly Detection System for Cloud Data Centres}

\author{Sakil Barbhuiya, Zafeirios Papazachos, Peter Kilpatrick and Dimitrios S. Nikolopoulos
\IEEEcompsocitemizethanks{\IEEEcompsocthanksitem S. Barbhuiya, Z. Papazachos, P. Kilpatrick, and D. S. Nikolopoulos are with 
the School of Electronics, Electrical Engineering and Computer Science, Queen's University Belfast, Belfast, United Kingdom.
Email: \{sakil.barbhuiya, z.papazachos, p.kilpatrick, d.nikolopoulos\}@qub.ac.uk}}

\IEEEtitleabstractindextext{
\begin{abstract}
Cybersecurity attacks in Cloud data centres are increasing alongside the growth of the Cloud services market. 
Existing research proposes a number of anomaly detection systems for detecting such attacks. 
However, these systems encounter a number of challenges, specifically due to the unknown behaviour of the attacks and the occurrence of genuine Cloud workload spikes, which must be distinguished from attacks. 
In this paper, we discuss these challenges and investigate the issues with the existing Cloud anomaly detection approaches. Then, we propose a Real-time Anomaly Detection System (RADS) for Cloud data centres, which uses a one class classification algorithm and a window-based time series analysis to address the challenges. 
Specifically, RADS can detect VM-level anomalies occurring due to DDoS and cryptomining attacks.
We evaluate the performance of RADS by running lab-based experiments and by using real-world Cloud workload traces. 
Evaluation results demonstrate that RADS can achieve 90-95\% accuracy with a low false positive rate of 0-3\%.
The results further reveal that RADS experiences fewer false positives when using its window-based time series analysis in comparison to using state-of-the-art average or entropy based analysis.

\end{abstract}

}

\maketitle

\IEEEdisplaynontitleabstractindextext



\section{Introduction}
\label{sec:introduction}
\noindent Cloud computing services are becoming ever more popular. According to a report\footnote{http://www.gartner.com/newsroom/id/3354117} from Gartner, Infrastructure as a Service (IaaS) has seen growth of more than 40\% in revenue every year since 2011. They also predict growth of more than 25\% every year through 2019 for IaaS. 
Such a growth in the Cloud services market has attracted various cybersecurity attackers to exploit vulnerabilities in the Cloud in order to gain personal benefit. 
Amongst the various cybersecurity attacks in the Cloud, DDoS and cryptomining attacks are growing sharply. 
In the Cloud, DDoS attacks typically attempt to overwhelm the Virtual Machine (VM) network by sending large amount of network packets from multiple hosts, so that the VM cannot serve its legitimate users' requests for various services such as web application, media streaming application, etc. 
Whereas, cryptomining attacks gain remote access to the VM in order to use its CPU computing power to perform cryptocurrency mining, which in turn interrupts legitimate users' computation on the VM. 
According to a report\footnote{https://www.cisco.com/c/en/us/solutions/collateral/service-provider/visual-networking-index-vni/vni-hyperconnectivity-wp.html} from Cisco, DDoS attacks with size greater than 1 Gbps increased by 172\% in 2016 (1.3 million attacks in 2016) and they predict that the attacks will increase to 3.1 million by 2021. Most recently, on February 28, 2018 the Github website was hit by the largest-ever DDoS attack\footnote{https://thehackernews.com/2018/03/biggest-ddos-attack-github.html} (1.35 Tbps).
Due to the rising value of cryptocurrency, cryptomining attacks increased six-fold during the period January-August 2017, as reported in Infosecurity magazine\footnote{https://www.infosecurity-magazine.com/news/ibm-cryptomining-attacks-increased/}. Most importantly, Cloud environments are very much vulnerable to cryptomining attacks due to the auto-scaling nature of the Cloud which allows the attackers to automatically spawn more VMs, i.e. more CPUs for the cryptomining task. This is evident from the recently identified cryptomining attack\footnote{https://www.coindesk.com/tesla-public-cloud-was-briefly-hijacked-by-crypto-miners/} on electric vehicle maker Tesla's Cloud environment. 

To successfully deny legitimate users access to the Cloud services and to perform cryptocurrency mining, both DDoS and cryptomining attacks significantly consume the network and the CPU of the Cloud VMs, respectively. This results in significant deviation in the normal network and CPU usage pattern of the VMs, which can be defined as VM-level anomaly. Hence, anomaly detection techniques can be used to identify DDoS and the cryptomining attacks in the Cloud. Researchers have proposed various anomaly detection techniques for Cloud which use machine learning or statistical approaches. The anomaly detection systems proposed in~\cite{ml_based:2012, Pandeeswari2016, ML_based_ids:2016} use supervised machine learning algorithms. These algorithms require both the ``normal" and the ``anomalous" behaviour traces to build the learning models, which can detect the anomalies. 
The algorithms may fail to detect anomalies arising due to unknown DDoS or cryptomining attacks, traces of which are not recorded by the learning models or which have very different patterns from the learned ``anomalous" patterns. To solve this problem researchers in~\cite{automated-detection:2016, UBL:2012, cloud-malware:2016} have proposed unsupervised learning and one class classification algorithms such as K-Means, Self Organising Map (SOM), and one class Support Vector Machine (SVM). 
These algorithms build the learning models by using the ``normal" behaviour traces. The models can identify anomalies by observing the deviation in the ``normal" behaviour pattern and as a result, these algorithms can successfully detect zero-day or unknown attacks.
Although the unsupervised learning and one class classification algorithms improve the accuracy of anomaly detection along with the ability to detect zero-day attacks, these algorithms may exhibit false positives arising due to workload spikes in a Cloud data centre. We can consider these spikes as genuine workload spikes which do not follow the ``normal" workload trend and their values are significantly higher than the other values in the workload data set. 
It is important to note that the genuine workload spikes persist only for a momentary period of time and this differentiates them from the anomalies (high utilisation values) due to DDoS and cryptomining attacks, which persist for a relatively long period of time.

In order to understand whether the VMs hosted in a real-world Cloud data centre experience workload spikes, we analysed real-world Cloud workload traces~\cite{workloadCCGRID:2015} collected from a Cloud data centre named Bitbrains\footnote{https://www.solvinity.com}. The traces contain seven performance metrics including CPU utilisation and network throughput of 1,750 VMs.
We performed a spike detection analysis by using the Interquartile Range (IQR\footnote{https://en.wikipedia.org/wiki/Interquartile\_range}) algorithm.
From the analysis of one month of the trace data from~\cite{workloadCCGRID:2015}, we observe that 84\% of VMs show spikes in their CPU utilisation at least once in the experimental month, whereas 95\% of VMs show spikes in their network traffic at least once in the same time period. 
From this finding we can anticipate that an anomaly detection system deployed in a large-scale Cloud data centre may generate frequent false positives due to the workload spikes. Receiving false positive alarms on a frequent basis is a major demerit of anomaly detection systems designed for the Cloud for a number of reasons: waste of operators' time as they engage in unnecessary investigations of the falsely raised alarms, unwanted interruption of users' applications while the operator tries to mitigate the anomaly without realising that the alarm is false, etc. 
This motivates a solution to remove false positives from the anomaly detection systems designed for the Cloud. 
Researchers in~\cite{automated-detection:2016}, \cite{UBL:2012}, \cite{cloud-malware:2016} consider window-based averaging on the raw data to reduce false positives. The works in~\cite{EbAT:2010} and \cite{entorpy_based_detection_2:2014} consider entropy-based anomaly detection which also reduces the number of false positives. However, these approaches may still generate false positives in certain scenarios for certain use cases, which we explain in the next section.

In this paper, we propose a Real-time Anomaly Detection System (RADS) for Cloud data centres, which can detect VM-level anomalies occurring due to unknown DDoS and cryptomining attacks. 
RADS uses a One Class Classification (OCC)~\cite{OCC:2008} based algorithm that learns the ``normal" pattern of CPU and network usage of each of the hosted VMs. 
The algorithm flags an anomaly whenever a VM's CPU or network usage pattern deviates significantly from its ``normal" pattern. 
To deal with the false positives, RADS combines average and standard deviation of the raw data in a window-based time series analysis.
Specifically, we make the following \textbf{key contributions} in this paper:
\begin{enumerate}[{(1)}]
\item We propose RADS for Cloud data centres which achieves high accuracy and low false positive rate in detecting VM-level anomalies occurring due to unknown DDoS and cryptomining attacks. RADS can operate in real-time, meaning that it can monitor each VM hosted in the Cloud data centre in real-time and detect the attacks as they appear inside the VMs.
\item We propose a novel training optimisation algorithm that decides the optimal amount of training data to be used for building the VM-specific classification models. This helps in achieving real-time dynamic training for RADS as opposed to offline static training which uses a fixed amount of training data. This is important considering the fact that in a Cloud data centre, the VMs host diverse workloads and a fixed amount of training data for all the VM-specific classification models may result in poor performance for RADS.
\item We evaluate the performance of RADS by running lab-based experiments in an OpenStack\footnote{https://www.openstack.org} based Cloud data centre. We emulate the DDoS and the cryptomining attacks by running microbenchmarks. Evaluation results show that RADS can detect VM-level anomalies with an accuracy of 90-95\% and a low false positive rate of 0-3\%. The results further reveal on average 34\% improvement in accuracy and 60\% improvement in false positive rate when RADS uses its window-based time series analysis instead of using the state-of-the-art average~\cite{automated-detection:2016}, \cite{UBL:2012}, \cite{cloud-malware:2016} or entropy~\cite{EbAT:2010}, \cite{entorpy_based_detection_2:2014} based analysis. 
\item We further validate the performance of RADS in terms of false positive rate by analysing real-world Cloud workload traces~\cite{workloadCCGRID:2015} collected from a Cloud data centre named Bitbrains. The analysis results demonstrate that RADS experiences fewer false positives when using its window-based time series analysis in comparison to using average~\cite{automated-detection:2016}, \cite{UBL:2012}, \cite{cloud-malware:2016} or entropy~\cite{EbAT:2010}, \cite{entorpy_based_detection_2:2014} based analysis.
\end{enumerate}

The remainder of the paper is organised as follows. Section~\ref{sec:problem_definition} defines the problems with the existing approaches in Cloud anomaly detection.
Section~\ref{sec:approach} demonstrates RADS window-based time series analysis. Section~\ref{sec:overview} and \ref{sec:framework} give an overview of RADS and discuss the RADS framework in detail, respectively. 
Section~\ref{sec:experiments} presents experimental results and discusses them. 
Section~\ref{sec:related_work} presents related work in Cloud anomaly detection which use different types of machine learning algorithms.
Finally, Section~\ref{sec:conclusions} concludes the paper. 

\section{Problem Definition}
\label{sec:problem_definition}
\noindent In order to detect cybersecurity attacks in a Cloud data centre, anomaly detection systems generally make the following assumptions:
\begin{itemize} 
\item \textbf{Assumption-1:} Resource utilisation of a VM follows some kind of ``normal” behaviour or trend which can be modelled by using machine learning or statistical approaches.
\item \textbf{Assumption-2:} Cybersecurity attacks such as DDoS and crytomining attacks consume VM resources significantly. This results in a deviation in the ``normal” trend of a VM's resource utilisation, which can be captured as anomalous by the machine learning or statistical approaches.
\end{itemize} 

Assumption-1 is very general and is considered by many anomaly detection systems like~\cite{automated-detection:2016}, \cite{UBL:2012}, \cite{cloud-malware:2016}, \cite{EbAT:2010}. 
Assumption-2 is experimentally demonstrated by~\cite{ddso_charater_2017} while they analysed real DDoS attack samples taken from the CAIDA\footnote{http://www.caida.org/data/passive/ddos-20070804\_dataset.xml} dataset. 
Recently, Radiflow\footnote{https://radiflow.com}, a cybersecurity solution provider, has discovered the first documented cryptomining attack on a SCADA network. According to Radiflow, cryptomining attacks cause high CPU and network bandwidth consumption. 
In our work, we also consider both these assumptions.


In most cases, the Cloud anomaly detection systems perform well under these assumptions. However, there are some cases where they may suffer from performance issues. In this paper we specifically consider the case where a Cloud anomaly detection system is using a linear classifier like K-Means, SVM, Naive Bayes, etc., and the VMs are exhibiting workload spikes. We explain this in the following example.

\begin{figure}[!h]
  \centering
   {\epsfig{file = 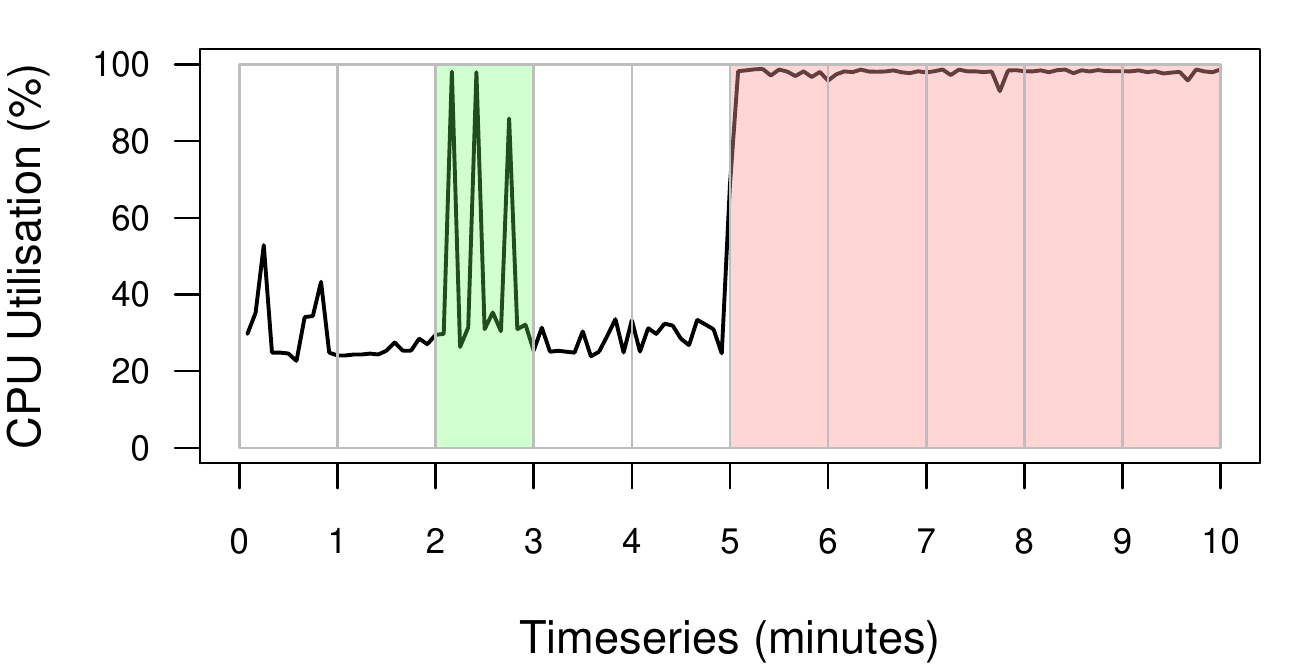, width = \columnwidth}}
   \caption{Time series of CPU utilisation while running the Graph Analytics application. Pink coloured sections represent the utilisation during the \textit{anomaly-period} and green coloured section represents the utilisation with workload spikes during the \textit{normal-period}}
  \label{fig:cpu_timeseries}
\end{figure}

\textit{Example Scenario:} We created an example scenario where a VM hosted in a Cloud data centre runs a Cloud application and at one stage the VM becomes compromised by a cryptomining attack that consumes its CPU to perform illicit cyrptocurrency mining. We built the Cloud data centre in our lab using OpenStack\footnote{https://www.openstack.org} (details of the set-up are available in Section~\ref{sec:experiments}) and executed a Graph Analytics workload (collected from CloudSuite\footnote{http://cloudsuite.ch}) as the Cloud application in one of the VMs hosted in our data centre. 
We emulated the cryptomining attack by running a CPU stress tool that consumes almost 100\% CPU of the VM. This emulation closely relates to real-world cryptomining attacks where the CPUs are consumed significantly to perform the mining. 
Considering this scenario, we performed 10 minutes of experiment to analyse the VM's CPU utilisation under different situations. 
We split the experimental period into two - (i) \textit{normal-period}: first 5 minutes, without any attack and (ii) \textit{anomaly-period}: last 5 minutes, under cryptomining attack. 
Furthermore, we injected some artificial workload spikes during the 3\textsuperscript{rd} minute by running the CPU stress tool for instantaneous periods of time consecutively. 
Figure~\ref{fig:cpu_timeseries} presents the time series graph of the CPU utilisation collected every 5 seconds. 

\begin{figure}[!h]
  \centering
   {\epsfig{file = 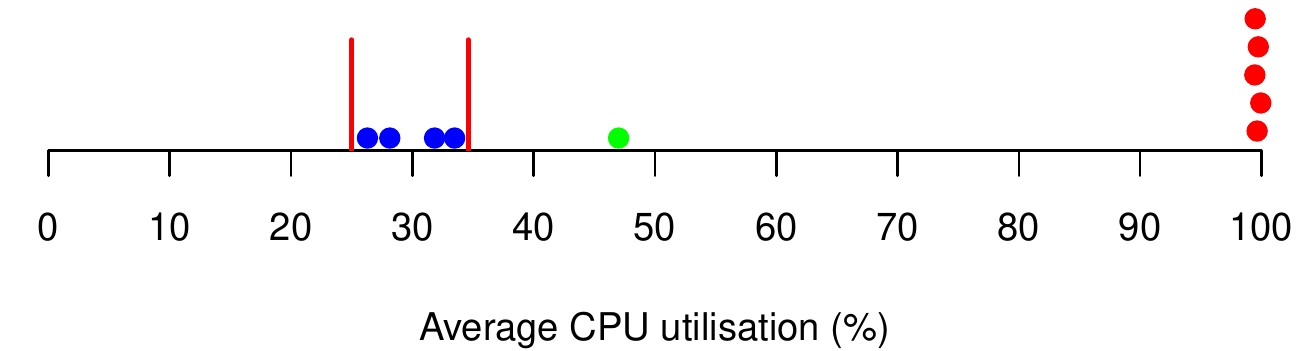, width = \columnwidth}}
     \hspace{0.9cm}
      {\epsfig{file = 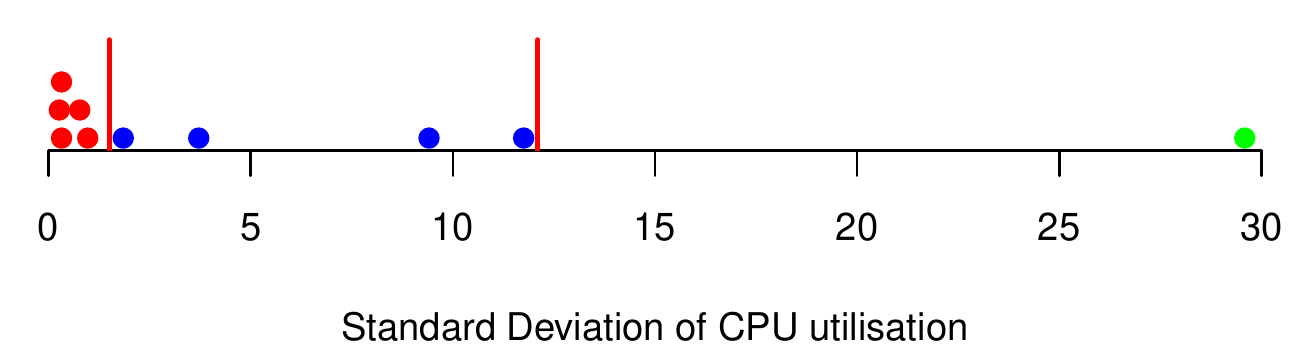, width = \columnwidth}}
        \hspace{0.9cm}
         {\epsfig{file = 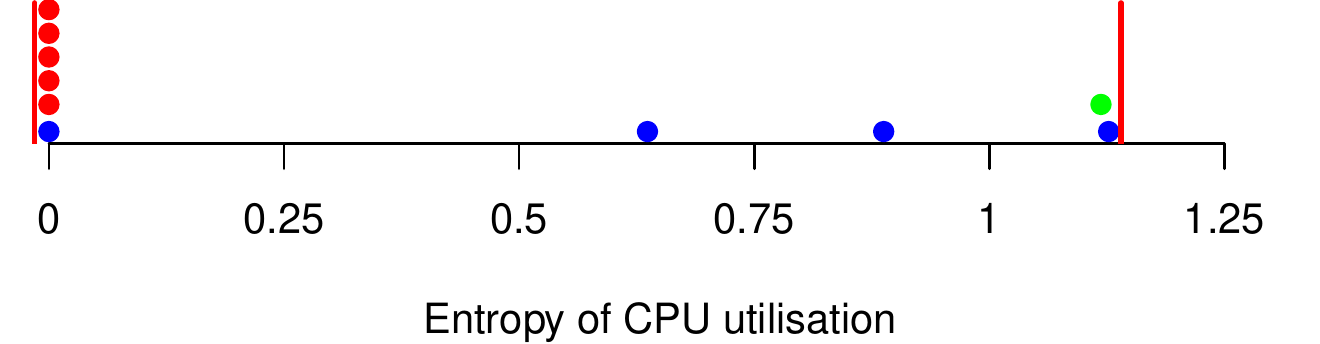, width = \columnwidth}}
   \caption{Average, standard deviation, and entropy of CPU utilisation}
  \label{fig:avg_sd_entropy}
\end{figure}

\textit{Window-based Time Series Analysis:} 
Anomaly detection systems which use linear classifiers, generally perform window-based time series analysis where the raw time series data are firstly distributed into a number of data bins with equal window size, and secondly, the average or entropy of the data is calculated in each bin. These averages or entropies collected from the bins form the time series data to be used in the anomaly detection systems.
To perform such an analysis, we grouped the CPU utilisation data points into 10 data bins, each with a window size of 1 minute (the grey coloured partitions of the time series in Figure~\ref{fig:cpu_timeseries}) and then calculated three statistical measurements (average, standard deviation, and entropy) of the CPU utilisation in each bin. We selected the window size to be 1 minute as we experimentally found that anything shorter than this does not help in reducing the noise from the CPU utilisation and anything longer than this does not capture the short-term CPU utilisation behaviour.  

For a discrete random variable $X$ with possible values $\big\{x_1,x_2...,x_n\big\}$ the entropy~\cite{entropy:2001} is calculated using Equation~\ref{entropy}. 
To prepare the data for the entropy calculation in each bin, we firstly normalise each of the raw data samples using Equation~\ref{normalisation} (normalised values are in the range $[0.0-1.0]$) and secondly, we decide to which amongst the following 10 smaller bins each normalised value belongs: 
$[0.0-0.1), [0.1-0.2), [0.2-0.3), [0.3-0.4), [0.4-0.5), [0.5-0.6), [0.6-0.7), [0.7-0.8), [0.8-0.9), [0.9-1.0]$. 
Finally, in each bin, we count the number of occurrences of the normalised values of the raw data samples in each smaller bin.
Thus, in Equation~\ref{entropy} we consider these numbers of occurrences as the values of the random variable $X$ in order to calculate the entropy.

We had 10 values (1 from each data bin) for each statistical measurement, which we present in Figure~\ref{fig:avg_sd_entropy} using 10 coloured dots along the x-axis. The 4 blue dots represent the measurements during the \textit{normal-period}, whereas the 5 red dots represent the measurements during the \textit{anomaly-period}. The green dot represents a measurement during the \textit{normal-period}, when the CPU encountered consecutive workload spikes (refers to the 3\textsuperscript{rd} minute in Figure~\ref{fig:cpu_timeseries}).

\begin{equation}
\label{entropy}
H(X) = -\sum\limits_{i=1}^n P(x_i)\log P(x_i)
\end{equation}
\begin{tabular}{l l}
     where &$P(x_i)$\;=\;probability mass function of $x_i$\\
                &$-logP(x_i)$\;=\;surprisal or self-information of $x_i$\\
\end{tabular}\\ 

\begin{equation}
\label{normalisation}
    X_{normalised} \quad = \quad   \frac{X - X_{min}}{X_{max} - X_{min}}
\end{equation}
\begin{tabular}{l l}
     where &$X_{normalised}$\;=\;normalised metric value\\
                &$X$\;=\;current metric value\\
                &$X_{min}$\;=\;minimum metric value in the raw data set \\
                &$X_{max}$\;=\;maximum metric value in the raw data set \\
\end{tabular}\\ \\ 

In the case of linear classifiers, the ``normal" data points are separated from the ``anomalous" data points by a hyperplane. 
In this analysis, we consider that the anomalies (red dots) and the genuine workload spikes (green dot) are appearing only during the testing or detection phase of the classifier. Therefore, for each statistical measurement, we drew two hyperplanes (the red lines) by considering the minimum and the maximum values of the ``normal" data points (blue dots); Figure~\ref{fig:avg_sd_entropy} presents this. We expected that the green dot (workload spikes) resides within the hyperplanes as they belong to the \textit{normal-period} and the red dots (anomalies) are clearly separable by the hyperplanes as they belong to the \textit{anomaly-period} of the experiment. 
From the figure we observe that, in the case of average, the blue dots are closely clustered and the red dots are clearly separable from them by the right hyperplane. However, the green dot representing the genuine workload spikes does not reside within the hyperplanes and indicates an anomaly.
In the case of standard deviation, although the blue dots are clustered together, the blue and the red dots are marginally separable by the left hyperplane, and importantly, the green dot does not reside within the hyperplanes and moves very far from both the blue and the red dots.
In the case of entropy, the blue and the red dots are not separable, although the green dot resides within the hyperplanes.
From these observations we identify the following problems for Cloud anomaly detection systems:
\begin{enumerate}[{(1)}] 
\item An average based linear classifier may identify the genuine workload spikes as anomalies and raise false positives.
\item Similar to average, a standard deviation based linear classifier may also raise false positives. Additionally, it may even fail to differentiate between normal behaviour and anomalies, which may raise false negatives resulting in low classification accuracy.  
\item An entropy based linear classifier may not raise false positives; but, similar to standard deviation, it may result in low classification accuracy due to failure in differentiating between normal behaviour and anomalies.
\end{enumerate}

\section{RADS Window-based Time-series Analysis}
\label{sec:approach}
\noindent 
\begin{figure}[!h]
  \centering
   {\epsfig{file = 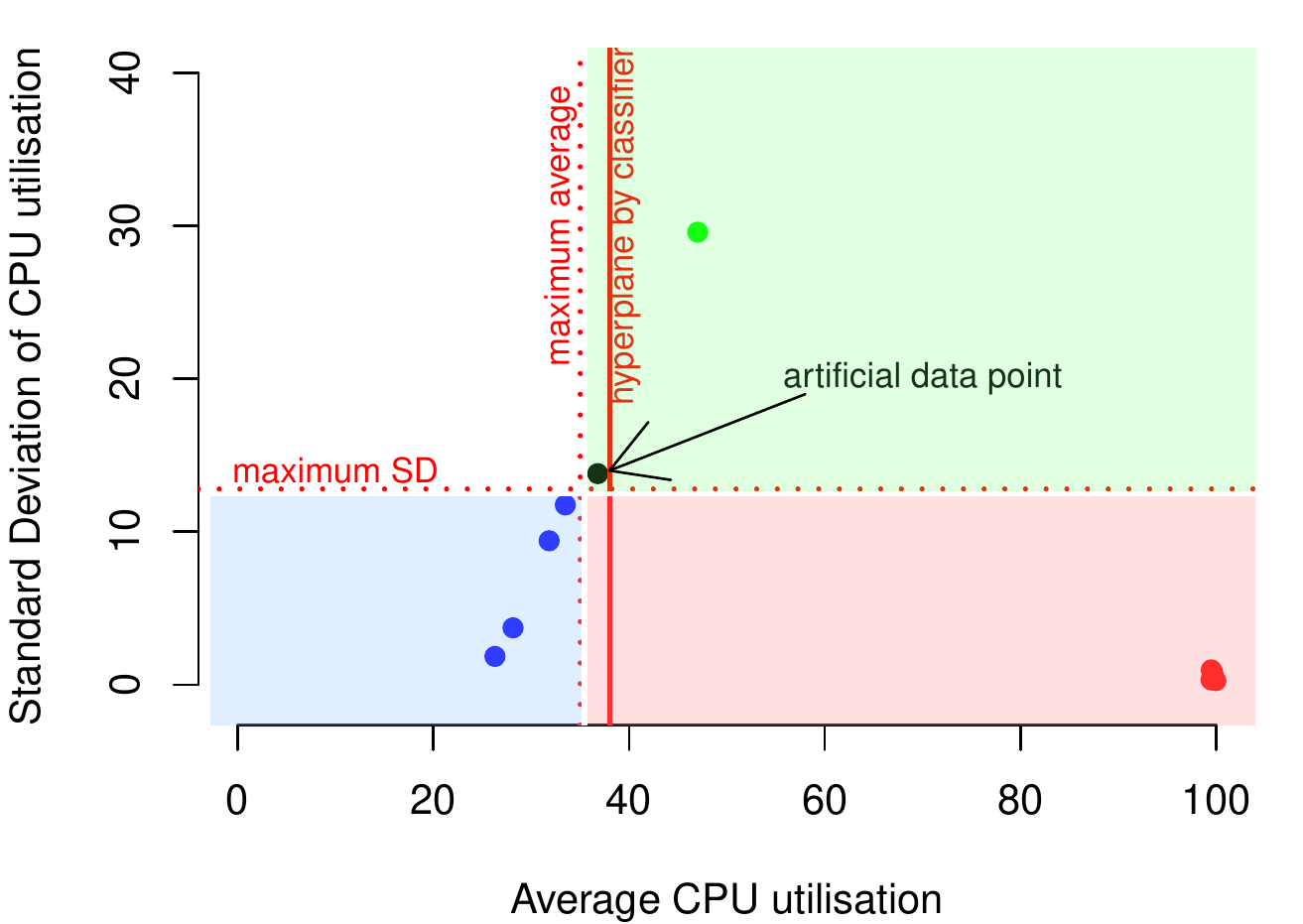, width = \columnwidth}}
   \caption{RADS approach: combining average and standard deviation}
  \label{fig:avg_sd_combined}
\end{figure}

\noindent In this section we explain RADS window-based time series analysis that resolves the problems identified in the previous section. 
Specifically, RADS combines average and standard deviation of the raw data in each time series window; and uses artificial data points that represent workload spikes. 

If we combine the average and standard deviation values generated from the experiment as discussed in the previous section, then we can represent them in a two-dimensional space as shown in Figure~\ref{fig:avg_sd_combined}. Similar to the previous section, blue, red, and green dots refer to the measurements during the normal, anomalous, and spike situations, respectively. From the figure we observe that the coloured dots can be classified into three classes if we draw the dotted red hyperplanes (horizontal and vertical) based on the maximum average on x-axis and maximum standard deviation (SD) on y-axis. Hence, this becomes a three class classification problem, where the classes can be labeled as: ``normal'' (blue coloured section) containing blue dots, ``anomaly'' (pink coloured section) containing red dots, and ``spike'' (green coloured section) containing green dot. However, we do not wish to go in that direction of classification as we assume that the samples for the ``anomaly'' class as well for the ``spike'' class are not available or known. 
RADS represents the green dot (``spike" class) with an artificial data point (black dot) which is a vector of the form: (max\_avg, max\_SD), where the max\_avg and the max\_SD are the maximum average and standard deviation of the blue dots (``normal'' class), respectively. This representation is based on the following assumptions: 

\begin{itemize} 
\item \textbf{Assumption-3:} Workload spikes exhibit average and standard deviation values higher than the maximum average and standard deviation values exhibited by ``normal" behaviour, respectively. That means that in Figure~\ref{fig:avg_sd_combined}, the assumption is that the green dot will never reside in the pink coloured section. 
\item \textbf{Assumption-4:} Anomalies exhibit standard deviation values lower than the maximum standard deviation value exhibited by ``normal" behaviour. That means that in Figure~\ref{fig:avg_sd_combined}, the assumption is that the red dots will never reside in the green coloured section. 
\end{itemize}  

We define the workload spikes as high utilisation values which persist only for a momentary period of time. Hence, in a time series window, the spikes will generate a high average value with a high standard deviation value and this will support Assumption-3.
We can support Assumption-4 with the fact that due to the nature of their attack, both DDoS and cryptomining attacks consume the resources significantly in a consistent manner without interrupt, whereas, resource consumption in a ``normal" behaviour is expected to have inconsistency and interruption. 

Thus, using the artificial data point RADS converts the three class classification problem into a two class classification problem where the classes are now: (i) ``positive'', which is composed of known ``normal" (blue dots) and unknown ``spike" (black dot) samples and (ii) ``negative'', which is composed of unknown ``anomaly" (red dots) samples. 
In Figure \ref{fig:avg_sd_combined} we can see that the two classes are clearly separable by a solid red hyperplane. Hence, a linear classifier can successfully differentiate between the two classes and produce high accuracy with low false positives. 

Similar to the CPU utilisation pattern deviation due to cryptomining attack, network traffic pattern deviates significantly due to DDoS attack. This is observed in~\cite{ddso_charater_2017} where they analysed DDoS attack samples taken from the CAIDA\footnote{http://www.caida.org/data/passive/ddos-20070804\_dataset.xml} dataset. Therefore, RADS analyses network traffic behaviour in the exactly the same manner as CPU utilisation behaviour analysis (discussed in this section) in order to detect VM-level anomalies occurring due to DDoS attack.

VMs may host varieties of applications in a Cloud data centre, some of which may be CPU intensive, some may be network intensive, and some may be both CPU and network intensive. Analysing both the CPU and network behaviour together makes the raw data points two dimensional, where in many cases one of the two parameters of the data points may generate steady time series data without any variance. In such cases, classification algorithms may suffer from the curse of dimensionality. We experimentally found this happening while executing two different Cloud applications (one CPU intensive and another network intensive) in our testbed. Therefore, RADS analyses the CPU and the network behaviour separately although it can perform both in parallel if required. 

\section{RADS Overview}
\label{sec:overview} 

\noindent In this section we discuss how RADS builds a linear classification model that differentiates between the ``positive" and the ``negative" classes as defined in the previous section, and we explain how RADS performs its real-time training and anomaly detection.

RADS aims to detect anomalies arising due to unknown DDoS and cryptomining attacks, traces of which are not previously recorded. Hence, we consider that the ``negative" class samples of the attacks are not available and RADS needs to build the classification model using the ``positive" class samples only. RADS achieves this by using the One Class Classification (OCC) algorithm that is proposed by Hempstalk et al. in~\cite{OCC:2008}. 
The algorithm first generates the artificial data (``negative" class) from a multi-variate normal distribution as estimated from the training data (``positive'' class) and, second, uses these artificial data as a second class in the construction of a binary class classification model, which is capable of classifying between the ``positive'' and the ``negative'' class. 
The classification is based on Bayes' Theorem\footnote{http://www.investopedia.com/terms/b/bayes-theorem.asp}.

\begin{figure}
  \centering
     \includegraphics[width=0.5\textwidth]{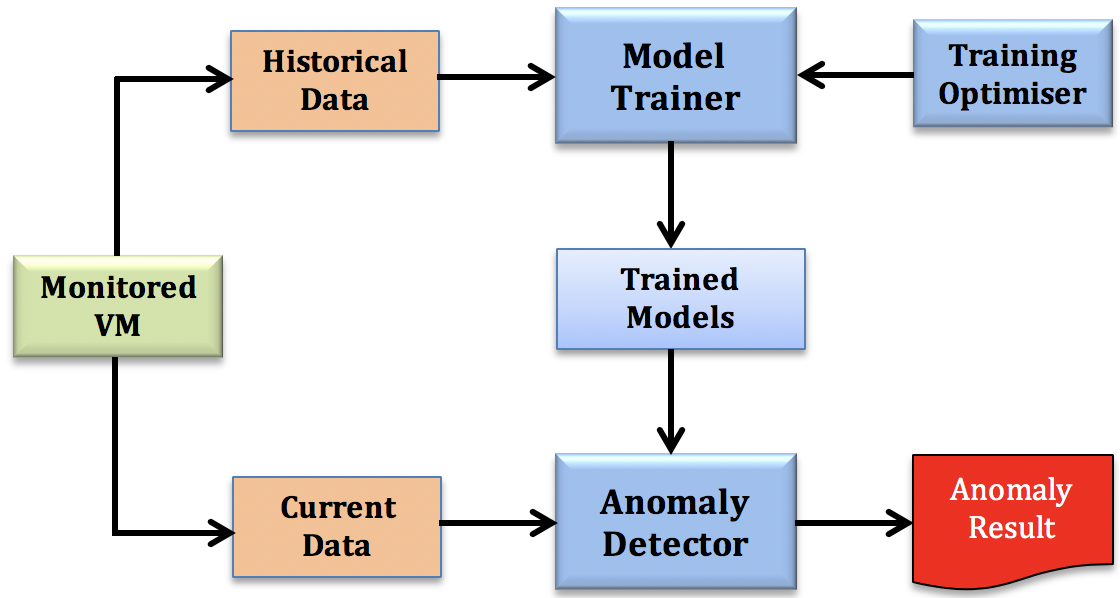}
   \caption{RADS overview}
  \label{fig:overview}
\end{figure}

Figure~\ref{fig:overview} depicts an overview of RADS. 
RADS runs the \textit{Model Trainer} to build or train the OCC models by using the ``positive" samples, i.e. the normal CPU utilisation or the network traffic data, which RADS collects from the hosted VMs in a Cloud data centre. Here it is important to note that RADS collects these training data assuming that the VMs are not affected by any DDoS or cryptomining attack. 
These training data are referred as the historical data as they are stored for a period of time. RADS uses the \textit{Training Optimiser} to decide the optimal amount of historical data to be used for the training of an OCC model. RADS runs the \textit{Anomaly Detector} to analyse the current data, i.e. the last one minute of CPU utilisation or the network traffic data by using the trained model. The model flags an anomaly whenever a VM's CPU or network usage pattern deviates significantly from its ``normal" pattern that is learned by the model. 
In both the training and the anomaly detection, RADS uses its window-based time series analysis as discussed in the previous section. 

\begin{algorithm}
\caption{Training Optimisation}
\label{raids_algorithm_training_optimiser}
\algnewcommand\algorithmicinput{\textbf{input:}}
\algnewcommand\algorithmicoutput{\textbf{output:}}
\algnewcommand\algorithmicabb{\textbf{abbreviation:}}
\algnewcommand\INPUT{\item[\algorithmicinput]}
\algnewcommand\OUTPUT{\item[\algorithmicoutput]}
\algnewcommand\ABB{\item[\algorithmicabb]}

\begin{algorithmic}[1]
\INPUT $SPT$ - \textit{Stability Period Threshold}
 \OUTPUT $TrainingStatus$ - first\_run/running/stopped/completed
 \ABB 
$ADR = Anomaly Detection Results$
\Statex
\For {\textbf{each} VM $vm_i$ \textbf{where} \textit{i=1,...,N}}
\If {$TrainingStatus_i != ``completed"$} 
\If {($TrainingStatus_i = ``first\_run"$ OR \textbf{ADR} contains ``anomaly")} 
\State RUN \textbf{Model Trainer}
\State $TrainingStatus_i = ``running"$
\State $stabilityPeriod_i = 0$
\State $break$
\Else 
\State $stabilityPeriod_i = stabilityPeriod_i + 5$
\EndIf
\If {$(stabilityPeriod_i = SPT)$}
\State $TrainingStatus_i = ``completed"$
\Else
\State $TrainingStatus_i = ``stopped"$
\EndIf
\State CLEAR $ADR\_File$
\EndIf
\EndFor
\end{algorithmic}
\end{algorithm}

\begin{algorithm}
\caption{Anomaly Detection}
\label{raids_algorithm_intrusion_detection}
\algnewcommand\algorithmicinput{\textbf{input:}}
\algnewcommand\algorithmicoutput{\textbf{output:}}
\algnewcommand\algorithmicabb{\textbf{abbreviation:}}
\algnewcommand\INPUT{\item[\algorithmicinput]}
\algnewcommand\OUTPUT{\item[\algorithmicoutput]}
\algnewcommand\ABB{\item[\algorithmicabb]}

\begin{algorithmic}[1]
\INPUT $CurrentData$ - last one minute of CPU utilisation or network traffic data for each VM; $M$ - set of trained OCC models (one for each VM)
 \OUTPUT $ADR$ - Anomaly Detection Result (one for each VM)
\Statex
\For {\textbf{each} VM $vm_i$ \textbf{where} \textit{i=1,...,N}}
\State $classificationResult_i=M_i.classify(CurrentData_i)$
 \If {$(classificationResult_i = ``positive")$}
\State $ADR_i= ``normal"$
\Else 
\State $ADR_i= ``anomaly"$
\EndIf
\EndFor
\end{algorithmic}
\end{algorithm}

\begin{figure*}
  \centering
     \includegraphics[width=1\textwidth]{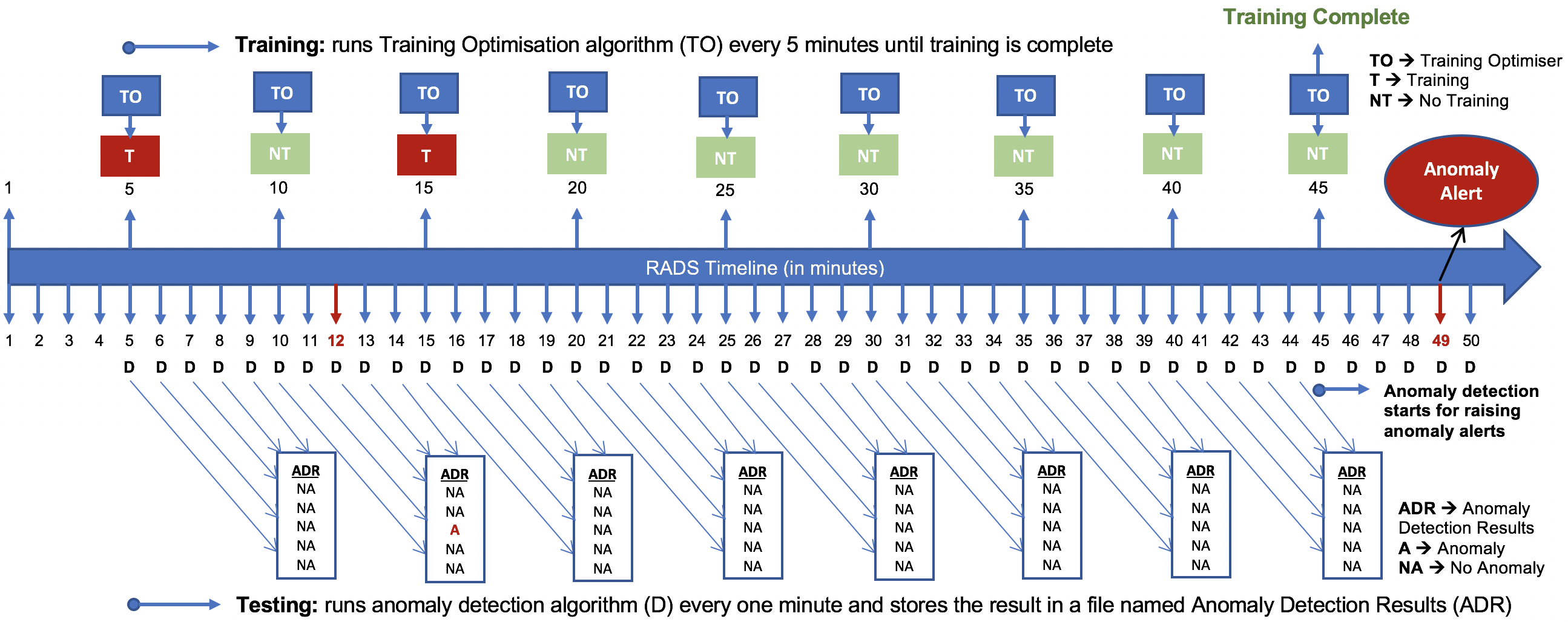}
   \caption{RADS  real-time training and anomaly detection}
  \label{fig:timeline}
\end{figure*}

We explain the real-time training and anomaly detection of RADS using a timeline as depicted in Figure~\ref{fig:timeline}. Specifically, we present the timeline of 50 minutes of RADS activity while performing training and detecting anomalies in real-time for a specific VM. We set up the behaviour of the VM artificially where the VM is behaving normally at all times except for minutes 12 and 49 where the VM experiences a genuine spike and an anomaly, respectively. For the first five minutes, RADS remain idle in order to accumulate data points to work with. At the end of 5\textsuperscript{th} minute, RADS starts its training which runs the training optimisation algorithm (TO) (Algorithm~\ref{raids_algorithm_training_optimiser}) every 5 minutes and starts its testing which runs the anomaly detection algorithm (D) (Algorithm~\ref{raids_algorithm_intrusion_detection}) every 1 minute. 
When the TO runs for the first time (at the end of 5\textsuperscript{th} minute), it performs the training to build the OCC model for the first time. In the later occasions, the TO evaluates the performance of the trained model by checking whether the model is identifying the VM's behaviour accurately without any false positive. 
This checking is performed by analysing the last five minutes of the anomaly detection results (ADR) obtained from D.
We assume that the VM is anomaly-free during the runtime of TO. 
Therefore, if the ADR contains ``anomaly" or ``A", that means that there is an anomaly falsely flagged by the trained model and the model needs to be trained again; in the timeline we can see that at the end of 15\textsuperscript{th} minute the training is performed again due to the occurrence of a false positive generated by the genuine spike at 12\textsuperscript{th} minute.
If the ADR does not contain ``anomaly" or ``A", that means that the model is correctly identifying the VM's behaviour and the model does not need further training; in the timeline we can see that at the end of 10\textsuperscript{th}, 20\textsuperscript{th}, 25\textsuperscript{th}, 30\textsuperscript{th}, 35\textsuperscript{th}, 40\textsuperscript{th}, and 45\textsuperscript{th} minutes the training is stopped. 

The period of time for which the model correctly identifies the ``normal" behaviour, is considered as the stability period for the model. The stability period is incremented with each correct identification by the model (e.g. in the timeline while moving from minute 15 to 45, the stability period is incremented to 30 minutes). Whenever the stability period reaches its threshold value, the TO declares that the training is complete; in the timeline we can see that at the end of 45\textsuperscript{th} minute as we set the threshold value to 30 minutes, which is based on the behaviour of the workloads executed in our testbed. 
However, the threshold for the stability period needs to be adjusted for different VMs based on their workload behaviour; a Cloud data centre may do this based on the type of the instances. 
Once the training is complete, RADS starts its anomaly detection for raising anomaly alerts; in the timeline we can see how RADS raises an anomaly alert at the 49\textsuperscript{th} minute due to the anomalous behaviour of the VM.

\section{RADS Framework}
\label{sec:framework} 

\begin{figure*}
  \centering
     \includegraphics[width=0.9\textwidth]{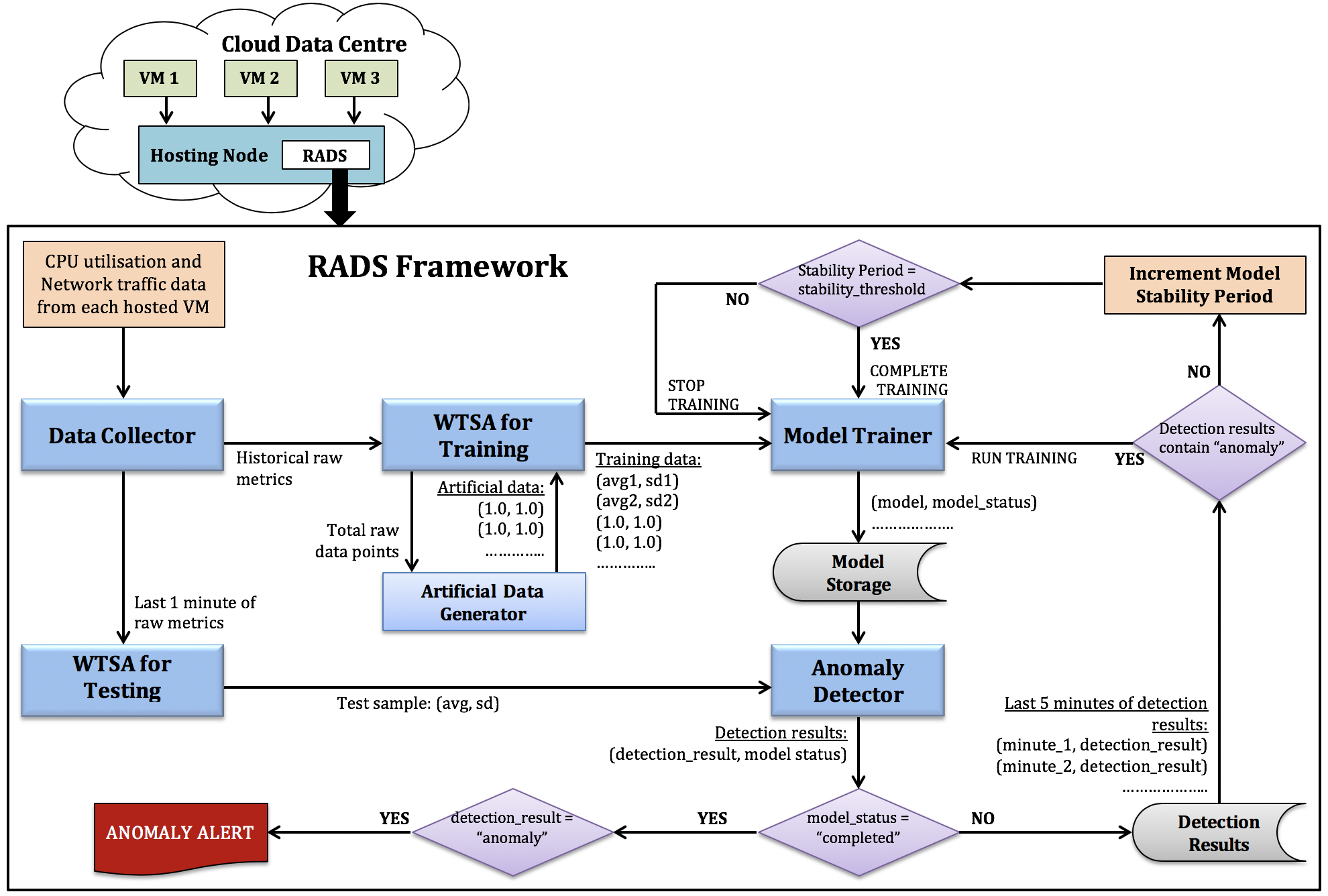}
   \caption{RADS framework}
  \label{fig:framework}
\end{figure*}

\noindent In this section we present the detailed framework of RADS. 
Figure~\ref{fig:framework} depicts the framework, which is designed to be implemented on each hosting node in a Cloud data centre locally, where it can monitor all the hosted VMs in order to detect the VM-level anomalies. 

\subsection{Data Collection}
\noindent RADS uses the \textit{Data Collector} module to collect the CPU utilisation and the network traffic (total size of network packets transmitted and received) metrics of each of the hosted VMs. The frequency of collecting these metrics is 5 seconds which allows capture of the CPU and network usage behaviour in a fine-grained manner.
The module runs virt-top\footnote{http://people.redhat.com/rjones/virt-top/} (a top-like utility for retrieving statistics of virtualised domains) on the hosting node for collecting the VM-level metrics. 

\begin{algorithm}
\caption{Window-based Time Series Analysis (WTSA) For Training}
\label{raids_algorithm_pre-processing_1}
\algnewcommand\algorithmicinput{\textbf{input:}}
\algnewcommand\algorithmicoutput{\textbf{output:}}
\algnewcommand\algorithmicabb{\textbf{abbreviation:}}
\algnewcommand\INPUT{\item[\algorithmicinput]}
\algnewcommand\OUTPUT{\item[\algorithmicoutput]}
\algnewcommand\ABB{\item[\algorithmicabb]}

\begin{algorithmic}[1]
\INPUT $Raw_{historical}$ - historical raw metrics of N VMs; $DW$ - distribution window = 1 minute
\OUTPUT $INN_{training}$ - set of normalised input instances for training; total N sets for N VMs
\ABB 
$avg = Average$; $sd = Standard Deviation$
\Statex
\For {\textbf{each} VM $vm_i$ \textbf{where} \textit{i=1,...,N}}
\State $dataBin(DB_{ij}) = Raw_{historical_i} / DW$ \thinspace \thinspace \thinspace \thinspace \textbf{where} \textit{j=1,...,B (total number of bins)}
\State $inputInstances(IN_i)=initiate()$
\For {\textbf{each} $DB_{ij}$}
\State $avg=DB_{ij}.getAvg()$
\State $sd=DB_{ij}.getSD()$
\State $inputInstance(in_j)=\big[avg, sd\big]$
\State $IN_i.addInstance(in_j)$
\State $IN_i.addClassLabel(``positive")$
\EndFor
\State $INN_{training_i} = IN_i.normalise() $ 
\For {\textbf{each} $DB_{ij}$}
\State $artificialInstance(art_j)=\big[1.0, 1.0\big]$ 
\State $INN_{training_i}.addInstance(art_j)$
\State $INN_{training_i}.addClassLabel(``positive")$
\EndFor 
\EndFor
\State \textbf{return} $INN_{training}$ \newline
\end{algorithmic}
\end{algorithm}

\subsection{Window-based Time Series Analysis (WTSA) For Training}
\noindent For each of the hosted VMs, the \textit{WTSA for Training} first takes all the historical raw metrics and distributes them into a number of data bins with equal window size of 1 minute, and second calculates the average (avg) and the standard deviation (sd) of the metrics in each bin. Thus, from each data bin, the module produces a vector: (avg, sd). 
In addition, the module generates artificial data points which represent the genuine workload spikes (see Section~\ref{sec:approach}). 
The number of artificial data points is equal to the total number of raw data points, which we have decided after evaluating the performance of RADS with varying number of artificial data points.
We represent each artificial data point as a vector: (1.0,1.0) which represents the maximum average and the maximum standard deviation values of the raw metrics as we consider the normalised values of the raw metrics. The normalisation is performed by using Equation~\ref{normalisation}. 
Finally, the module produces a series of vectors for use as training data by combining the vectors which are generated by performing the window-based processing of the raw metrics and the vectors which are generated artificially. 
We present the algorithm for this module in Algorithm~\ref{raids_algorithm_pre-processing_1}.

\subsection{Window-based Time Series Analysis (WTSA) For Testing}
\noindent The \textit{WTSA For Testing} module takes only the last one minute of raw metrics and calculates the average (avg) and the standard deviation (sd) of these metrics to produce the vector (avg, sd). This vector is considered as the test sample for detecting an anomaly. We present the algorithm for this module in Algorithm~\ref{raids_algorithm_pre-processing_2}. The algorithm normalises the \textit{avg} and \textit{sd} values before they are combined as a vector, as defined in Equation~\ref{normalisation}. Importantly, this normalisation is performed against the training data, where minimum and the maximum values are taken from the training data set. 
This is necessary in order to achieve the same normalisation for both the training and the testing data.

\subsection{Model Training}
\noindent RADS builds an OCC model for each hosted VM using the \textit{Model Trainer} module. The OCC models take the training data samples (generated by the \textit{WTSA for Training} module) as the input and learn the ``normal" CPU or network usage pattern of the VMs. 
The module stores the OCC models in the \textit{Model Storage}. 

\subsection{Anomaly Detection}
\noindent RADS detects the anomalies using the \textit{Anomaly Detector} module. For each VM, the module takes as input the test sample (generated by the \textit{WTSA for Testing} module) and the stored OCC model built for that VM. The module flags an ``anomaly" when there is a deviation in the VM's CPU or network usage pattern. 

We implemented the RADS modules using Java programming, which imports Apache Common Maths\footnote{http://commons.apache.org/proper/commons-math/} and Weka\footnote{http://www.cs.waikato.ac.nz/ml/weka/} libraries for performing statistical operations and One Class Classification (OCC).

\begin{algorithm}
\caption{Window-based Time Series Analysis (WTSA) For Testing}
\label{raids_algorithm_pre-processing_2}
\algnewcommand\algorithmicinput{\textbf{input:}}
\algnewcommand\algorithmicoutput{\textbf{output:}}
\algnewcommand\algorithmicabb{\textbf{abbreviation:}}
\algnewcommand\INPUT{\item[\algorithmicinput]}
\algnewcommand\OUTPUT{\item[\algorithmicoutput]}
\algnewcommand\ABB{\item[\algorithmicabb]}

\begin{algorithmic}[1]
\INPUT $Raw_{current}$ - last one minute of raw metrics of N VMs
\OUTPUT $INN_{testing}$ - normalised input instances for testing; total N instances for N VMs
\ABB 
$avg = Average$; $sd = Standard Deviation$
\Statex
\For {\textbf{each} VM $vm_i$ \textbf{where} \textit{i=1,...,N}}
\State $inputInstances(IN_i)=initiate()$
\State $avg_{normalised}=Raw_{current_i}.getAvg().normalise() $
\State $sd_{normalised}=Raw_{current_i}.getSD().normalise() $
\If {$ (avg_{normalised} > 1.0)$ AND $(sd_{normalised} > 1.0)$}
\State $inputInstance(in)=\big[1.0,1.0\big]$
\Else
\State $inputInstance(in)=\big[avg_{normalised}, sd_{normalised}\big]$
\EndIf
\State $IN_i.addInstance(in)$
\State $INN_{testing_i} = IN_i $ 
\EndFor
\State \textbf{return} $INN_{testing}$ \newline
\end{algorithmic}
\end{algorithm}

%

\section{Evaluation}
\label{sec:experiments}
\noindent We performed a number of experiments to evaluate the performance of RADS. The experiments can be classified into: lab-based and real-world. The lab-based experiments were performed in an OpenStack\footnote{https://www.openstack.org} based Cloud data centre, which hosted two representative Cloud applications drawn from the CloudSuite\footnote{http://cloudsuite.ch} benchmark suite. The real-world experiments were carried out on the real-world workload traces~\cite{workloadCCGRID:2015} collected from a Cloud data centre named Bitbrains\footnote{https://www.solvinity.com}.
In this section we present the results from these experiments and discuss them. 
Specifically, we attempt to answer the following research questions: 
\begin{enumerate}[{(1)}]
\item Can RADS accurately detect Cloud anomalies occurring due to DDoS and cryptomining attacks in real-time? 
\item Can RADS window-based time series analysis outperform the state-of-the-art average and entropy based analyses in terms of accuracy and false positive rate? 
\item Can RADS be used as a lightweight tool in terms of consuming minimal computing resources and processing time in a Cloud data centre? 
\item Does RADS maintain its performance in terms of removing false positives while analysing real-world Cloud workload traces?
\end{enumerate}

\subsection{Lab-based Experiments}
\label{sec:realtime_analysis}
\noindent In this section we evaluate the performance of RADS in detecting DDoS and cryptomining attacks in our lab-based Cloud data centre. Also, we evaluate the efficiency of RADS in terms of the system resources that it consumes and the time it takes while performing real-time training of the classification models and testing of new samples to detect the anomalies.
In particular, in this section we attempt to answer research questions 1, 2, and 3.

\textbf{Testbed:} Our testbed is an OpenStack based Cloud data centre which consists of four compute nodes. Each compute node is a Dell PowerEdge R420 server that runs CentOS 6.6 and has 24 cores, 2-way hyper-threaded, clocked at 2.20 GHz with 12GB DRAM clocked at 1600 MHz. The nodes include two 7.2K RPM hard drives with 1TB of SATA in RAID 0 and a single 1GBE port. KVM is the default hypervisor of the nodes. 

\begin{figure}[!h]
  \vspace{-0.2cm}
  \centering
   {\epsfig{file = 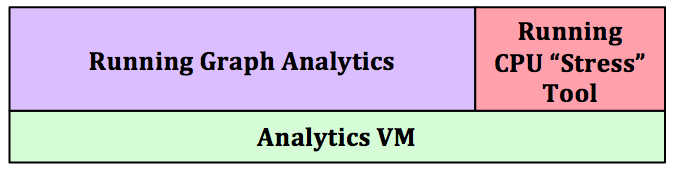, width = 0.7\columnwidth}}
      \caption{Experimental set-up for Graph Analytics application}
  \label{fig:testbed_analytics}
  \vspace{-0.1cm}
\end{figure}

\begin{figure}[!h]
  \vspace{-0.2cm}
  \centering
   {\epsfig{file = 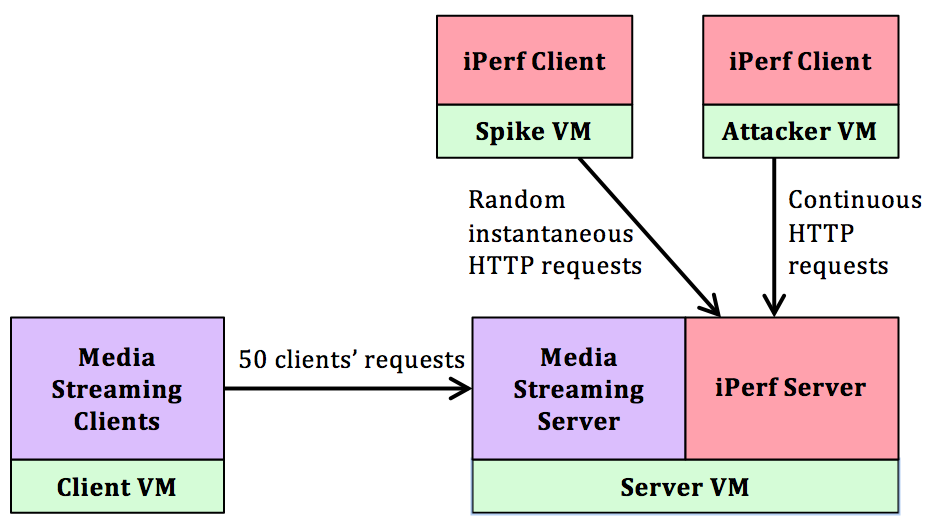, width = \columnwidth}}
         \caption{Experimental set-up for Media Streaming application}
  \label{fig:testbed_media}
  \vspace{-0.1cm}
\end{figure}

\textbf{Experimental Set-up:} We hosted two Cloud applications in our testbed: Graph Analytics (representing CPU intensive Cloud applications) and Media Streaming (representing network intensive Cloud applications). The experimental set-up for the Graph Analytics is depicted in Figure~\ref{fig:testbed_analytics} and the experimental set-up for the Media Streaming is depicted in Figure~\ref{fig:testbed_media}.
The Graph Analytics application performs PageRank on a Twitter dataset using the Spark\footnote{http://spark.apache.org} framework. We deployed the application on a dedicated ``Analytics VM" with the configuration: 8GB of RAM and 4 cores of CPU. Under ``normal" conditions we executed the application using only 1 core of CPU.
The Media Streaming application runs a streaming server using the Nginx\footnote{https://github.com/nginx/nginxweb} server, which hosts videos of various lengths and qualities.  
The clients send requests to the hosted videos to create realistic media streaming behaviour.  
We deployed the server on a dedicated ``Server VM" with the configuration: 4GB of RAM and 4 cores of CPU and the clients on a dedicated ``Client VM" with the same configuration.
In our experiment, we portrayed ``normal" media streaming behaviour by running 50 clients in the ``Client VM" with ``ShortHi" configuration which requests videos with high bandwidth of 790 Kbps.

\textit{Emulated attacks}: We emulated the DDoS and the cryptomining attacks targeting the VMs running the Media Streaming server and the Graph Analytics application, respectively. 
Cryptomining attack was emulated by running the ``stress"\footnote{https://people.seas.harvard.edu/\~apw/stress/} tool on the ``Analytics VM" (see Figure~\ref{fig:testbed_analytics}). The ``stress" tool is a simple workload generator, which can impose a configurable amount of CPU, memory, I/O, and disk stress on the system. 
The DDoS attack was emulated by sending continuous HTTP requests from an ``Attacker VM" to the ``Server VM" (see Figure~\ref{fig:testbed_media}) with the help of the iPerf\footnote{https://iperf.fr} tool (``Server VM" as iPerf server and ``Attacker VM" as iPerf client).

\textit{Emulated workload spikes}: For the Graph Analytics application, a workload spike was generated by running the ``stress" tool in the ``Analytics VM" for a short period of time (5 seconds). For Media Streaming application, a workload spike was generated by sending HTTP requests for instantaneous period of time (5 seconds) from the ``Attacker VM" to the ``Sever VM" (see Figure~\ref{fig:testbed_media}) with the help of the iPerf tool (``Server VM" as iPerf server and ``Spike VM" as iPerf client).


\textbf{Performance Metrics:} 
We use a number of standard performance metrics such as precision, recall, accuracy (F1 score), and false positive rate (FPR) to measure the performance of RADS. 
RADS reacts with an anomaly alarm whenever it classifies a test sample as ``anomalous", otherwise RADS does not react. 
In our experiments, we declare: (a) False Positives (FP) when RADS raises an alarm but there is no ``anomaly", (b) False Negatives (FN) when RADS fails to raise an alarm but there exists an ``anomaly", (c) True Positives (TP) when RADS raises an alarm and there exists an ``anomaly", (d) True Negatives (TN) when RADS does not raise an alarm and there is no ``anomaly". We define the performance metrics in Equations \ref{eq2}-\ref{eq5}.

\begin{equation}\label{eq2}
    Precision \quad = \quad   \frac{TP}{TP+FP}
\end{equation}
\begin{equation}\label{eq3}
    Recall \quad = \quad   \frac{TP}{TP+FN}
\end{equation}
\begin{equation}\label{eq4}
    Accuracy (F1 score) \quad = \quad   2\times\frac{Precision \times Recall}{Precision+Recall}
\end{equation}
\begin{equation}\label{eq5}
    FPR \quad = \quad   \frac{FP}{FP+TN}
\end{equation}

Precision gives us the measure of how many of the positive classifications (anomaly alarms) are correct, whereas the recall gives us the measure of RADS's ability to correctly identify an ``anomaly". 
However, precision and recall alone cannot judge the performance of RADS. Therefore, we use Accuracy (F1 score) which gives us the harmonic mean of precision and recall.


\newcolumntype{L}[1]{>{\raggedright\arraybackslash}p{#1}}
\newcolumntype{C}[1]{>{\centering\arraybackslash}p{#1}}
\newcolumntype{R}[1]{>{\raggedleft\arraybackslash}p{#1}}
\begin{table*}[t]
\caption{Anomaly detection results of RADS under different time series analyses}
\label{tab:test_results} 
\centering
\begin{tabular}{C{2.2cm}C{4.2cm}C{1.7cm}C{1.7cm}C{1.7cm}C{1.5cm}C{1.6cm}} 
  \toprule
  Monitored VM & Time Series Analysis & Training Time (minutes) & \textit{Attack Test} Result & \textit{Spike Test} Result 
  & Accuracy (F1 Score) & False Positive Rate (FPR)\\ 
    \bottomrule
   & & &\begin{tabular}{C{0.5cm}C{0.5cm}}  TP & FN \\  \end{tabular} & \begin{tabular}{C{0.5cm}C{0.5cm}}  FP & TN \\  \end{tabular} \\
    \bottomrule 
        \begin{tabular}{C{2cm}} Graph Analytics VM \\ \end{tabular} 
        & \begin{tabular}{C{4.0cm}} Average \\ Entropy \\ Average \& Standard Deviation \\ \end{tabular} 
        & \begin{tabular}{C{1.2cm}} 45 \\ 105 \\ 130 \end{tabular}
         &\begin{tabular}{C{0.5cm}C{0.5cm}} 10 & 0 \\ 0 & 10 \\ 9 & 1 \end{tabular}
         &\begin{tabular}{C{0.5cm}C{0.5cm}} 6 & 24 \\ 2 & 28 \\ 1 & 29 \end{tabular} 
         & \begin{tabular}{C{1.0cm}} 0.77 \\ 0.00 \\ 0.90 \end{tabular}
        & \begin{tabular}{C{1.0cm}} 0.20 \\ 0.07 \\ 0.03 \end{tabular} \\

  \hline
    \begin{tabular}{C{2cm}} Media Streaming Server VM \\ \end{tabular} 
       & \begin{tabular}{C{4.0cm}} Average \\ Entropy \\ Average \& Standard Deviation \\  \end{tabular} 
       & \begin{tabular}{C{1.2cm}} 70 \\ 20 \\ 35 \end{tabular}
       &\begin{tabular}{C{0.5cm}C{0.5cm}} 10 & 0 \\ 10 & 0 \\ 9 & 1 \end{tabular}
       &\begin{tabular}{C{0.5cm}C{0.5cm}} 8 & 22 \\ 0 & 30 \\ 0 & 30 \end{tabular} 
        & \begin{tabular}{C{1.0cm}} 0.71 \\ 1.00 \\ 0.95 \end{tabular}
       & \begin{tabular}{C{1.0cm}} 0.27 \\ 0.00 \\ 0.00 \end{tabular} \\
    \hline
      
\end{tabular}
\end{table*}

\textbf{Anomaly Detection Performance of RADS:}
To evaluate the anomaly detection performance of RADS we carried out two tests: (i) \textit{Attack Test -} during which we emulated the DDoS attack (targeting the VM running the Media Streaming server) or the cryptomining attack (targeting the VM running the Graph Analytics application) continuously for 10 minutes; and (ii) \textit{Spike Test -} during which we emulated workload spikes for 10 times in a random manner in a time period of 30 minutes; there is no emulated attack during this test.
During both the tests, we executed the RADS \textit{Anomaly Detector} module on the hosting node. 
For both the tests, the module used the same OCC models which were trained by the RADS \textit{Model Trainer} module. 

\begin{figure}[!h]
  \centering
      {\epsfig{file = 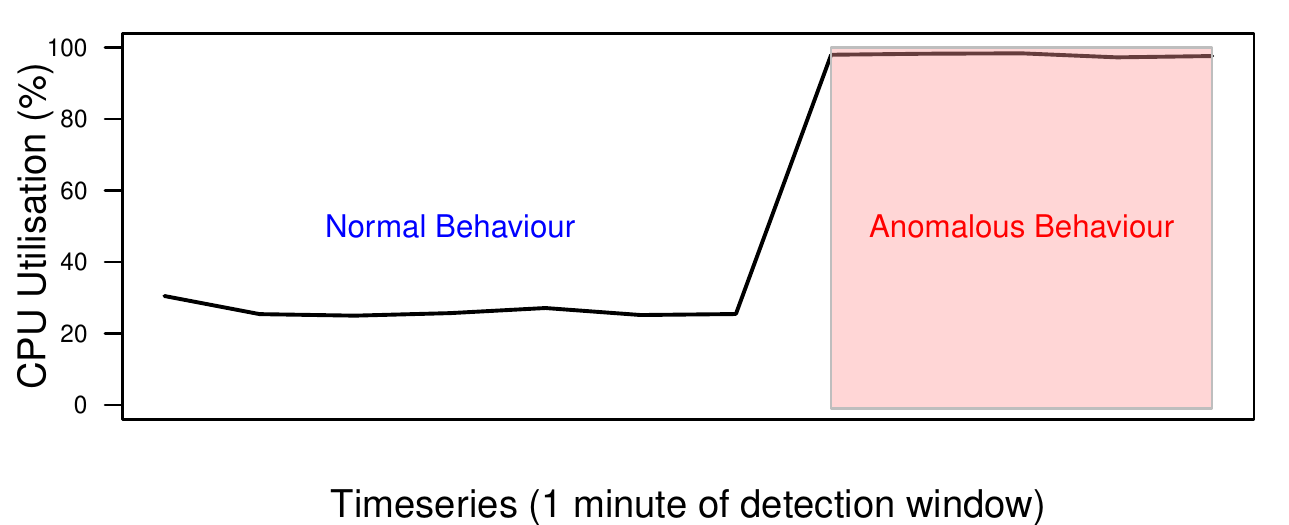, width = 0.9\columnwidth}}
   \caption{The first minute of detection window which includes both the ``normal" and the ``anomalous" behaviour}
  \label{fig:one_minute_timeseries}
\end{figure}

Table~\ref{tab:test_results} presents the test results under different time series analyses.
From the results we observe 
that the average based analysis achieves the maximum number of true positives (total 20), but on the other hand, this approach raises the maximum number of false positives (total 14).
Entropy raises only 2 false positives in the case of the Graph Analytics VM and no false positives in the case of the Media Streaming server VM, but the problem with the entropy based approach is its poor performance in detecting the attacks (10 false negatives in the case of the Graph Analytics VM).
RADS window-based time series analysis, which uses a combination of the average and the standard deviation, successfully detects the attacks with total 18 true positives. 
We have observed that the false negatives (1 for each of the monitored VMs) arise during the first minute of the \textit{Attack Test}, when the ``anomalous" behaviour due to attack does not occupy the whole 1 minute of the detection window; the time series of the detection window becomes similar to the one depicted in Figure~\ref{fig:one_minute_timeseries}. A test sample generated from such a detection window can be represented as: (high\_average, high\_standard\_deviation), which is wrongly classified as ``normal" by RADS as it considers the short-term ``anomalous" behaviour in the detection window as a genuine workload spike.
However, these false negatives are trivial due to the fact that they appear only during the first minute of the attack; and DDoS or cryptomining attacks require a considerable amount of time (at least a few minutes) before they become harmful.  

In order to get a better insight into the anomaly detection performance of RADS, we calculated the accuracy (F1 score) and the false positive rate (FPR) (see Table~\ref{tab:test_results}) using Equations \ref{eq4} and \ref{eq5}, respectively.  
These performance metrics answer the research questions 1 and 2 as follows:
\begin{enumerate}[{(a)}]
\item RADS can detect Cloud anomalies occurring due to DDoS and cryptomining attacks in real-time with an accuracy (F1 Score) of 90-95\% and a low false positive rate of 0-3\%.
\item RADS achieves on average 34\% improvement in accuracy and 60\% improvement in false positive rate while using its window-based time series analysis instead of using the state-of-the-art average or entropy based analysis. 
\end{enumerate}

\textbf{Efficiency of RADS:}
To evaluate the efficiency of RADS we carried out experiments while scaling up the number of hosted VMs from 2 to 10. Although such scaling of VMs does not represent a real Cloud data centre, we attempt to extract some information on RADS efficiency under VM scaled up situations. 

\begin{figure}[!h]
  \vspace{-0.2cm}
  \centering
   {\epsfig{file = 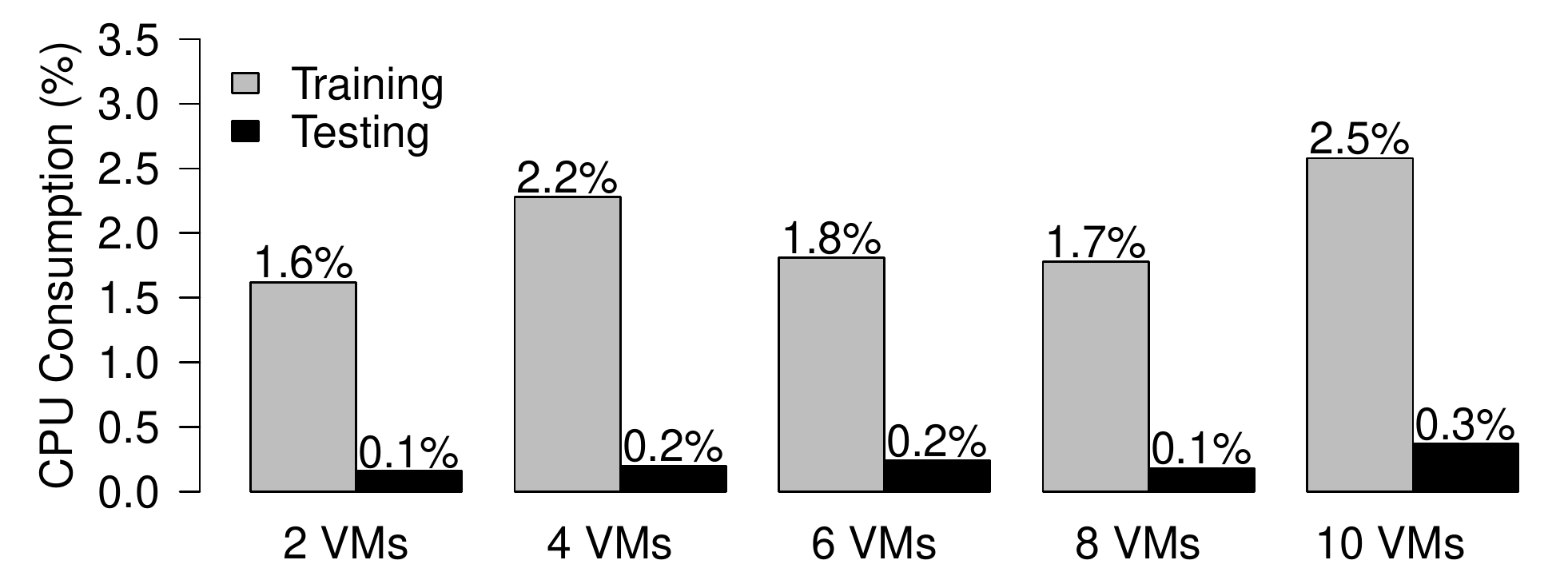, width = 0.9\columnwidth}}
   \caption{Hosting node CPU consumption by RADS}
  \label{fig:overhead_cpu}
  \vspace{-0.1cm}
\end{figure}

\begin{figure}[!h]
  \vspace{-0.2cm}
  \centering
   {\epsfig{file = 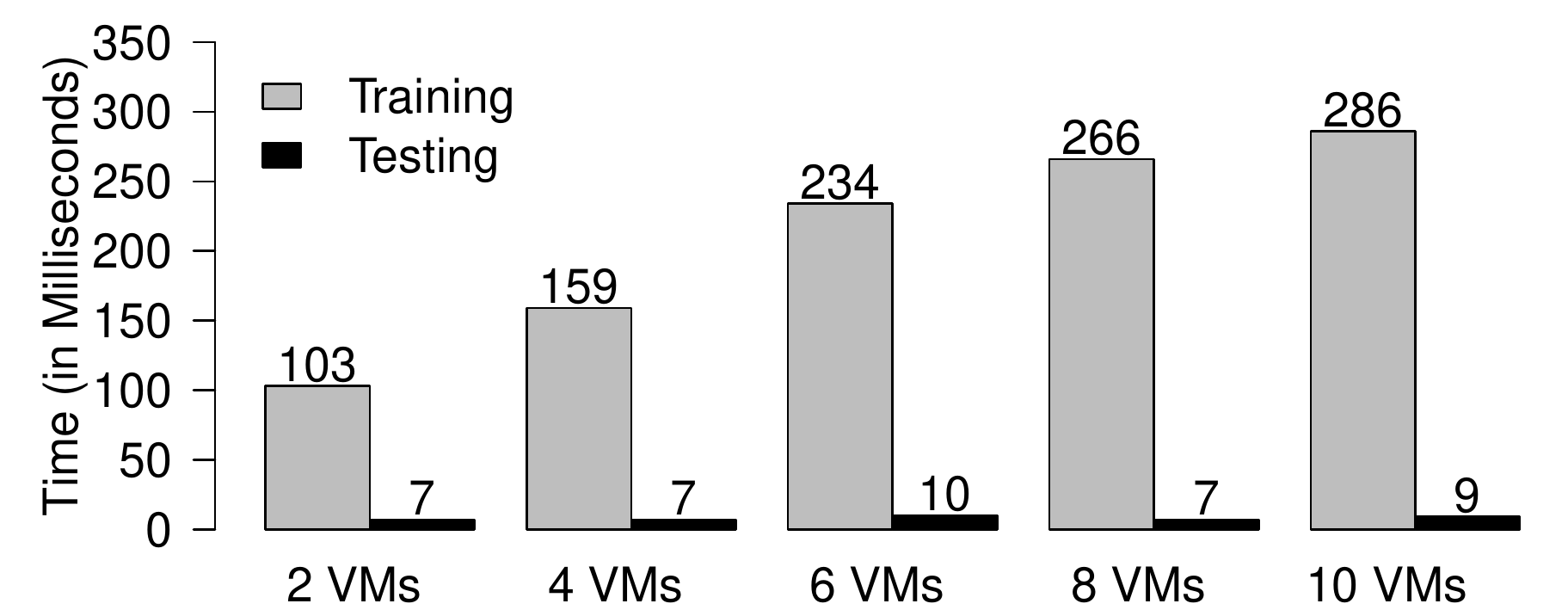, width = 0.9\columnwidth}}
   \caption{Training and testing time required by RADS}
  \label{fig:overhead_time}
  \vspace{-0.1cm}
\end{figure}

\textit{Computation cost of RADS:} We measured the computation cost of RADS in terms of its CPU consumption on the hosting node. 
The bar plots in Figure~\ref{fig:overhead_cpu} show the CPU consumed by RADS while performing the training and the testing (anomaly detection).
From the plots we find that for training, the CPU consumption remains very low (in the range 1.6\% to 2.5\%) and it does not increase much with the scaling up of the number of VMs, whereas for testing, the CPU consumption remains consistently negligible. 

\textit{Processing time of RADS}: The bar plots in Figure~\ref{fig:overhead_time} show the processing time that RADS took while performing the training and the testing. From the plots we observe that RADS took milliseconds in finishing the training and the testing tasks. The testing time is much lower than the training time. 
The training time increases with the scaling up of the number of VMs, but the testing time remains almost constant. 

In answering the research question 3, we can summarise that RADS can be used as a lightweight tool in terms of consuming minimal hosting node CPU and processing time in a Cloud data centre. However, the processing time required for training increases with the scaling up of the number of hosted VMs. This may lead to a RADS efficiency issue in the case where there are hundreds or thousands of hosted VMs and when the duration of the training increases to few hours or days. In future, we will attempt to explore this issue and address it with shared-memory or multithreaded programming solutions such as OpenMP, MPI, Phoenix++, etc.
%

\subsection{Real-world Experiments}
\label{sec:offline_analysis}
\noindent In this section we evaluate the performance of RADS in terms of false positive rate by analysing real-world Cloud workload traces. Specifically, in this section we attempt to answer research question 4.

\textbf{Trace Description:}
We selected the traces collected from a Cloud data centre named Bitbrains\footnote{https://www.solvinity.com} as analysed in~\cite{workloadCCGRID:2015}. 
Bitbrains specialises in managed hosting and business computation for enterprises such as banks, credit card operators, insurers, etc. 
The traces contain seven performance metrics including CPU utilisation and network throughput of 1,750 VMs. The metrics are sampled every 5 minutes. The traces were collected between July and September 2013 in two trace directories: (i) \textit{fastStorage} which consists of 1,250 VMs that are connected to fast storage area network (SAN) storage devices and (ii) \textit{Rnd} which consists of 500 VMs that are either connected to the fast SAN devices or to much slower Network Attached Storage (NAS) devices. \textit{fastStorage} contains one month of trace (August, 2013), whereas \textit{Rnd} contains three months of trace (July-September, 2013).
\textbf{Preparation of Traces:}
\label{sec:preparaiton_of_traces}
In order to use the traces from~\cite{workloadCCGRID:2015} for evaluating the performance of RADS, we made the following selection process:
\begin{enumerate}[{(1)}]
\item We selected only the traces from the month August, 2013 for which both the traces (\textit{fastStorage} and \textit{Rnd}) were available. 
\item We further selected the first three days of traces, making the assumption that the Cloud applications or workloads running inside the VMs are consistent throughout the experimental period. 
\item Out of the three days of traces, we selected the first two days (14:40, 12 August to 14:40, 14 August 2013) of traces for training and and the third day (14:40, 14 August to 14:40, 15 August 2013) of traces for testing. 
\item We performed spike detection analysis on the traces using the Interquartile Range (IQR\footnote{https://en.wikipedia.org/wiki/Interquartile\_range}) algorithm and selected traces only from VMs which flagged spikes. This is because we intend to evaluate the performance of RADS in terms of dealing with the genuine workload spikes observed in the traces. 
\item Finally, for CPU utilisation analysis, we selected traces from VMs which have CPU utilisation greater than 10\%, and for network throughput analysis we selected the traces from VMs with network throughput greater than 100KB/s. This is done in order to select traces from active VMs which are running a decent amount of workload. 
\end{enumerate}

Following the above selection process, we chose the traces from 82 VMs for CPU utilisation analysis and traces from 212 VMs for network throughput analysis.

\begin{figure}[!h]
  \centering
   {\epsfig{file = 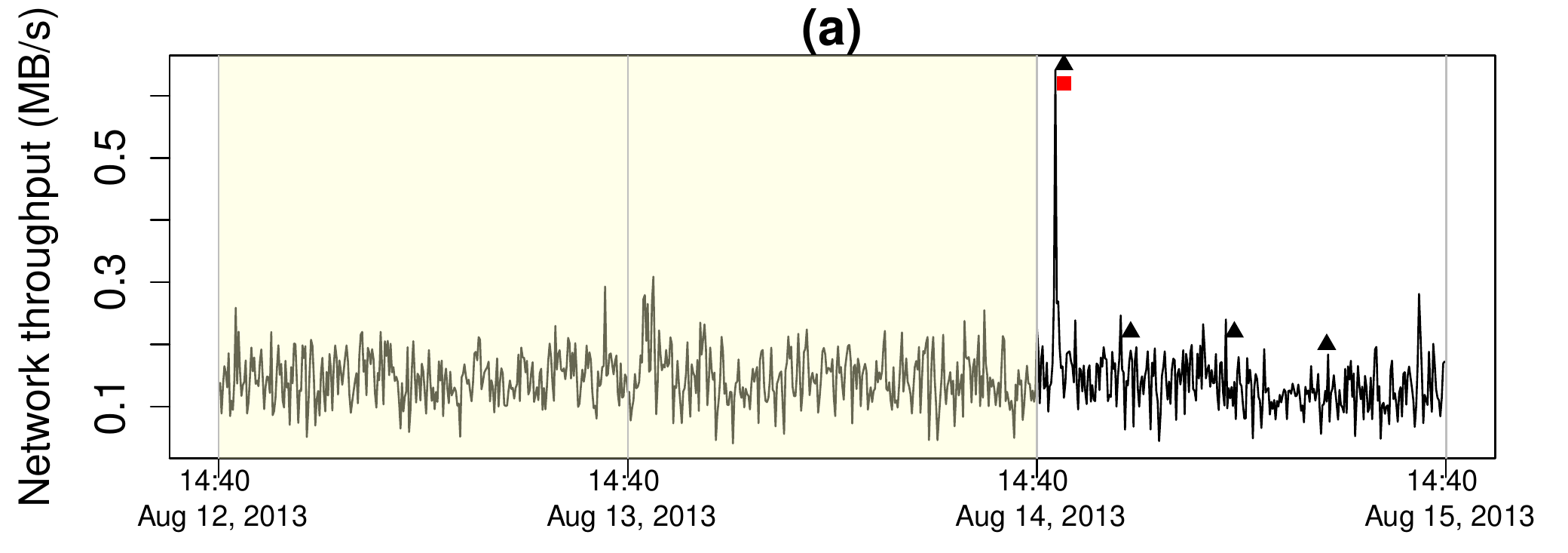, width = \columnwidth}}
    {\epsfig{file = 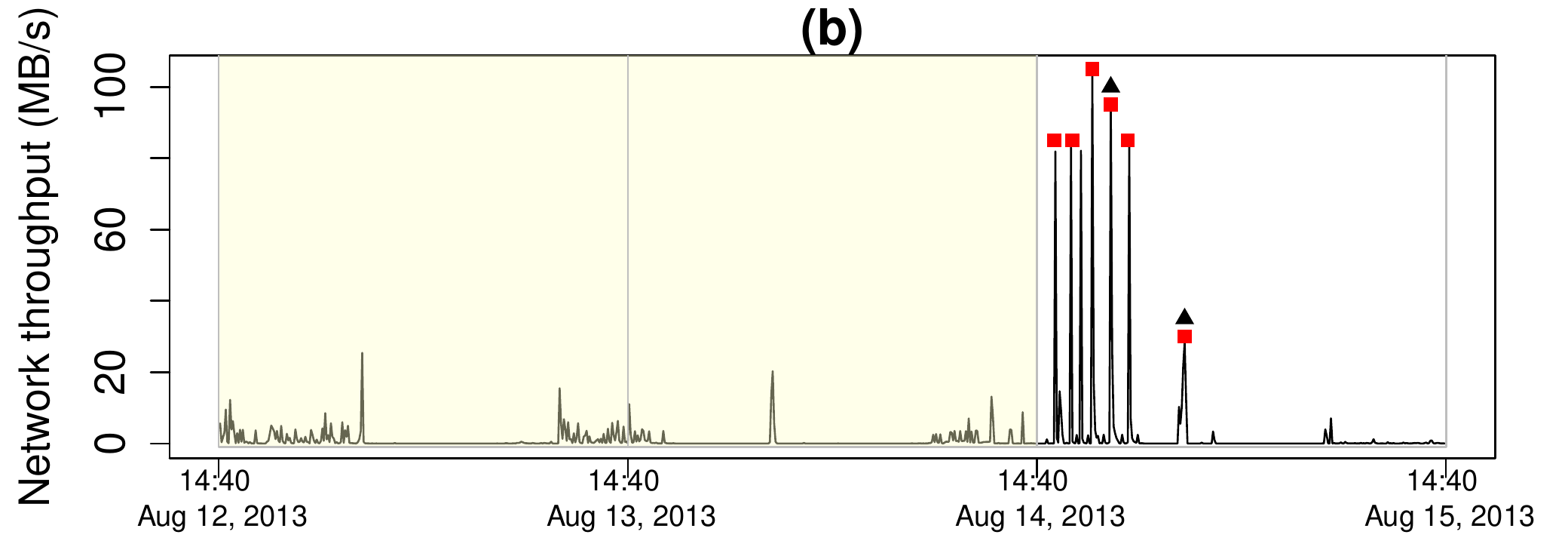, width = \columnwidth}}
      {\epsfig{file = 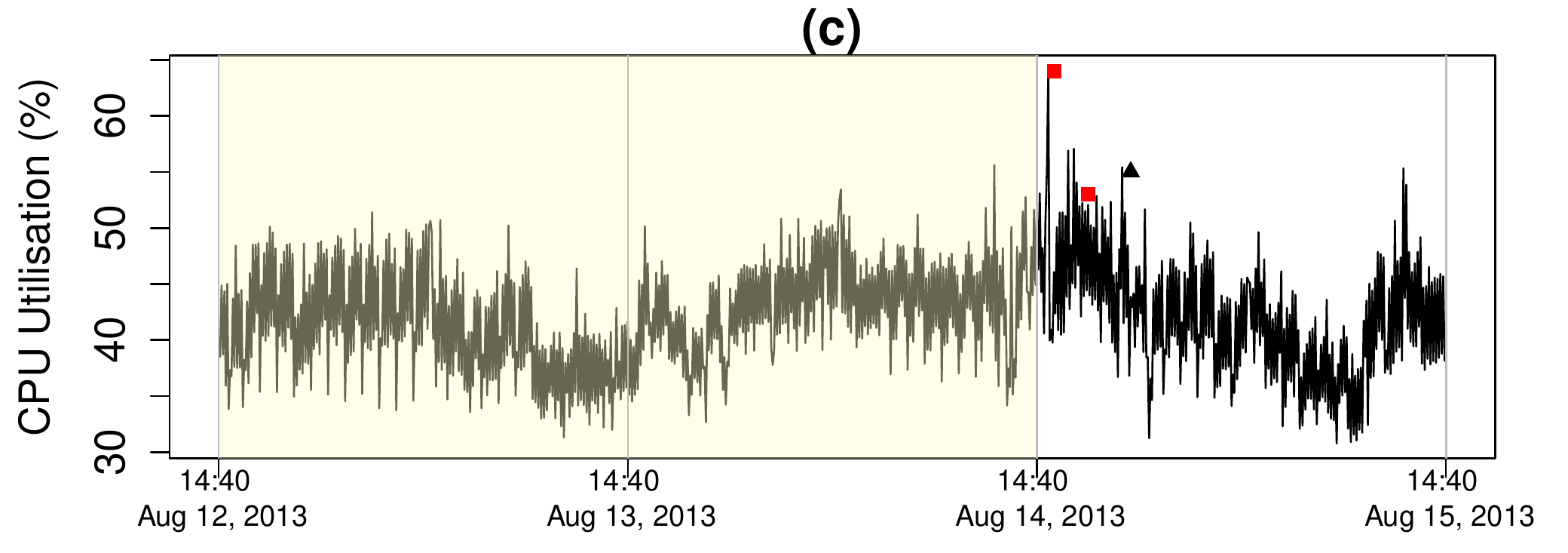, width = \columnwidth}}
         {\epsfig{file = 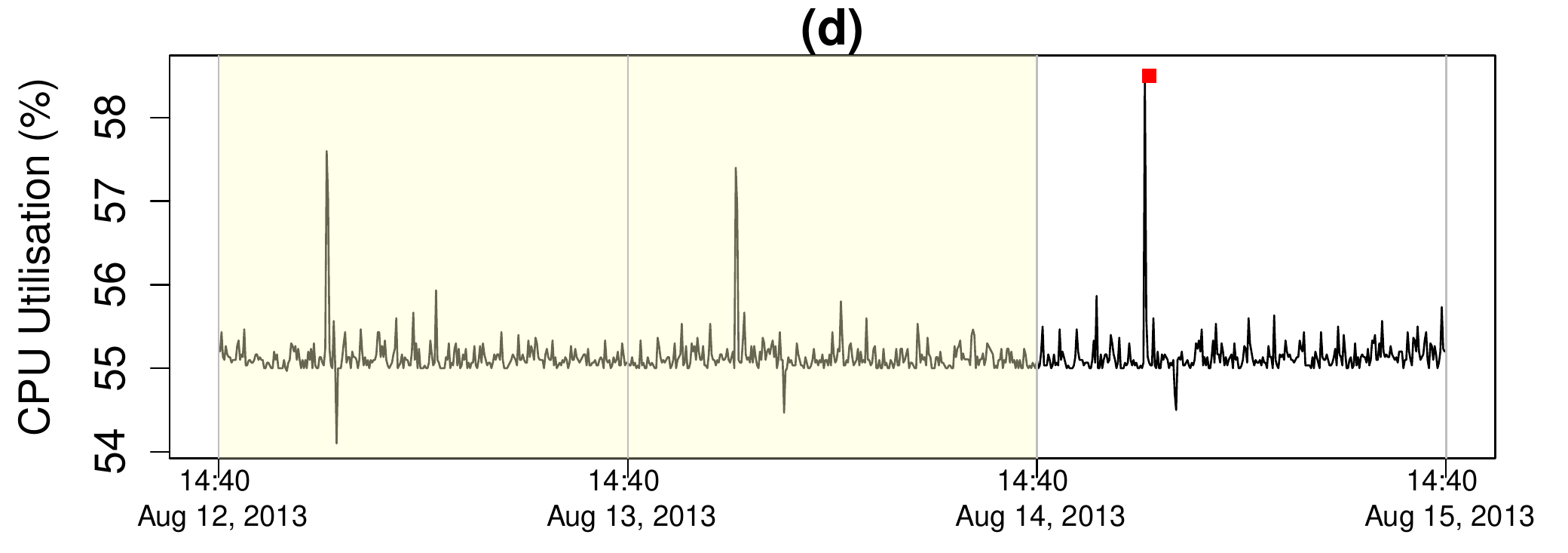, width = \columnwidth}}

                      \caption{RADS analysis of different Cloud workload behaviour observed in the traces collected from~\cite{workloadCCGRID:2015}: (a) VM-941 from fastStorage trace, (b) VM-357 from fastStorage trace, (c) VM-980 from fastStorage trace, (d) VM-306 from Rnd trace. The yellow coloured section represents the training period and the section without colour represents the testing period. The coloured shapes represent ``anomaly" alarms raised by RADS while using different time series analyses: red square and black triangle are for the average and the entropy based analysis, respectively. }

  \label{fig:offline_timeseries}
\end{figure}

\textbf{Performance of RADS Under Different Cloud Workloads:}
Out of the selected traces, we chose the traces from a range of VMs exhibiting varying workload behaviour as presented in Figure~\ref{fig:offline_timeseries} using the time series graphs. 
The time series graphs reveal how RADS performs under different Cloud workload behaviour while using different time series analyses. We summarise the observations from these graphs as follows:

\begin{enumerate}[{(a)}]
\item In both cases where the workload experiences consistently fluctuating behaviour (Figure~\ref{fig:offline_timeseries}(a)) and irregular behaviour (Figure~\ref{fig:offline_timeseries}(c)), RADS successfully classifies the genuine workload spikes as ``normal" while using its window-based time series analysis. But, while using the average or the entropy based analysis, RADS fails to classify the genuine workload spikes as ``normal" and raises false ``anomaly" alarms. 
\item In both the cases where the workload experiences significant genuine workload spikes (Figures~\ref{fig:offline_timeseries}(a) and (b)) and insignificant genuine workload spikes (Figure~\ref{fig:offline_timeseries}(c)), RADS successfully classifies them as ``normal" while using its window-based time series analysis.  But, while using the average or the entropy based analysis RADS fails to classify them as ``normal" and raises false ``anomaly" alarms.
\item While using its window-based time series analysis, RADS continues its successful classification of genuine workload spikes as ``normal" even when the workload experiences genuine workload spikes during the training period (Figure~\ref{fig:offline_timeseries}(d)). However, using the average based analysis RADS fails again to classify the genuine workload spikes as ``normal" and raises false ``anomaly" alarm. 
\end{enumerate}

%
%


\begin{figure}[!h]
  \centering
   {\epsfig{file = 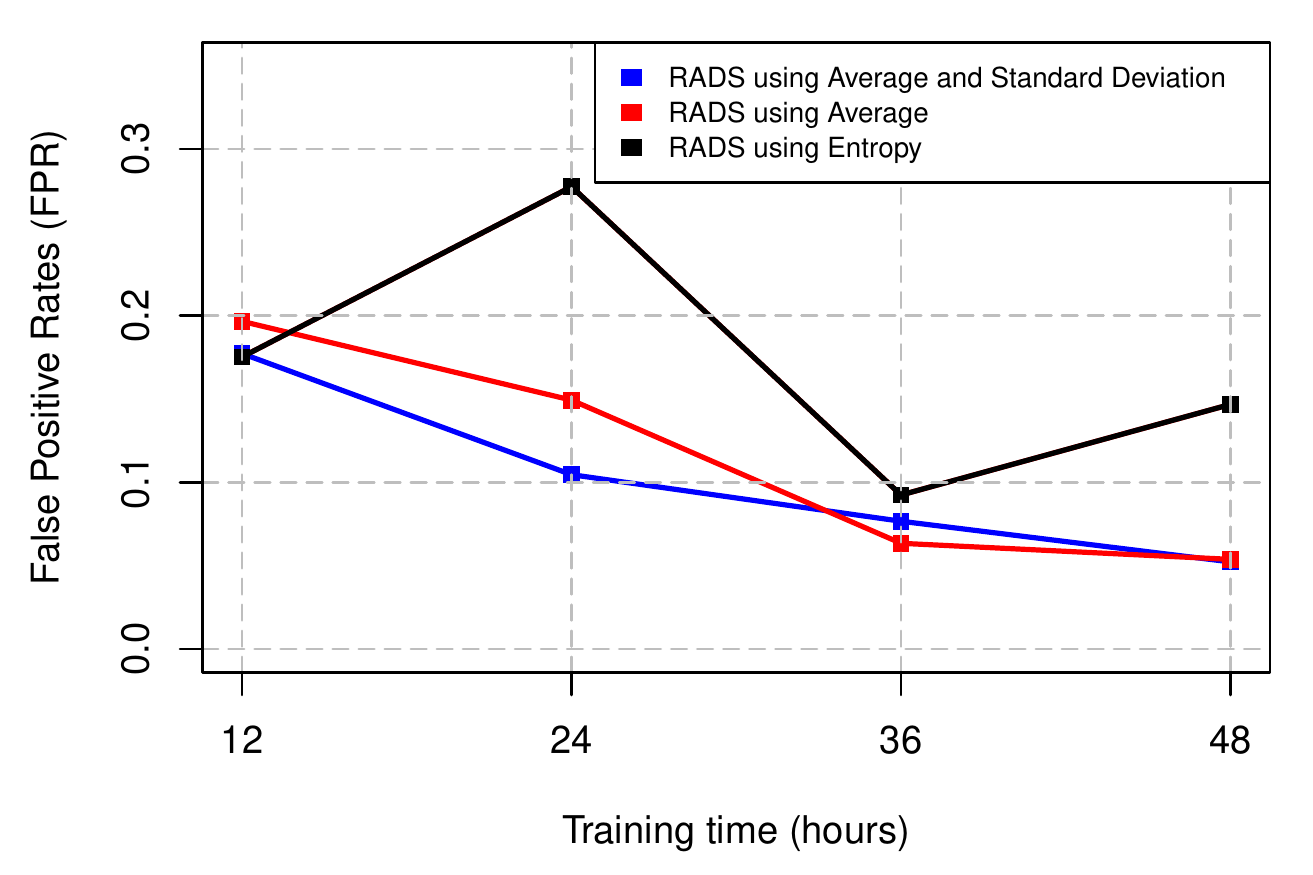, width = 0.9\columnwidth}}
      {\epsfig{file = 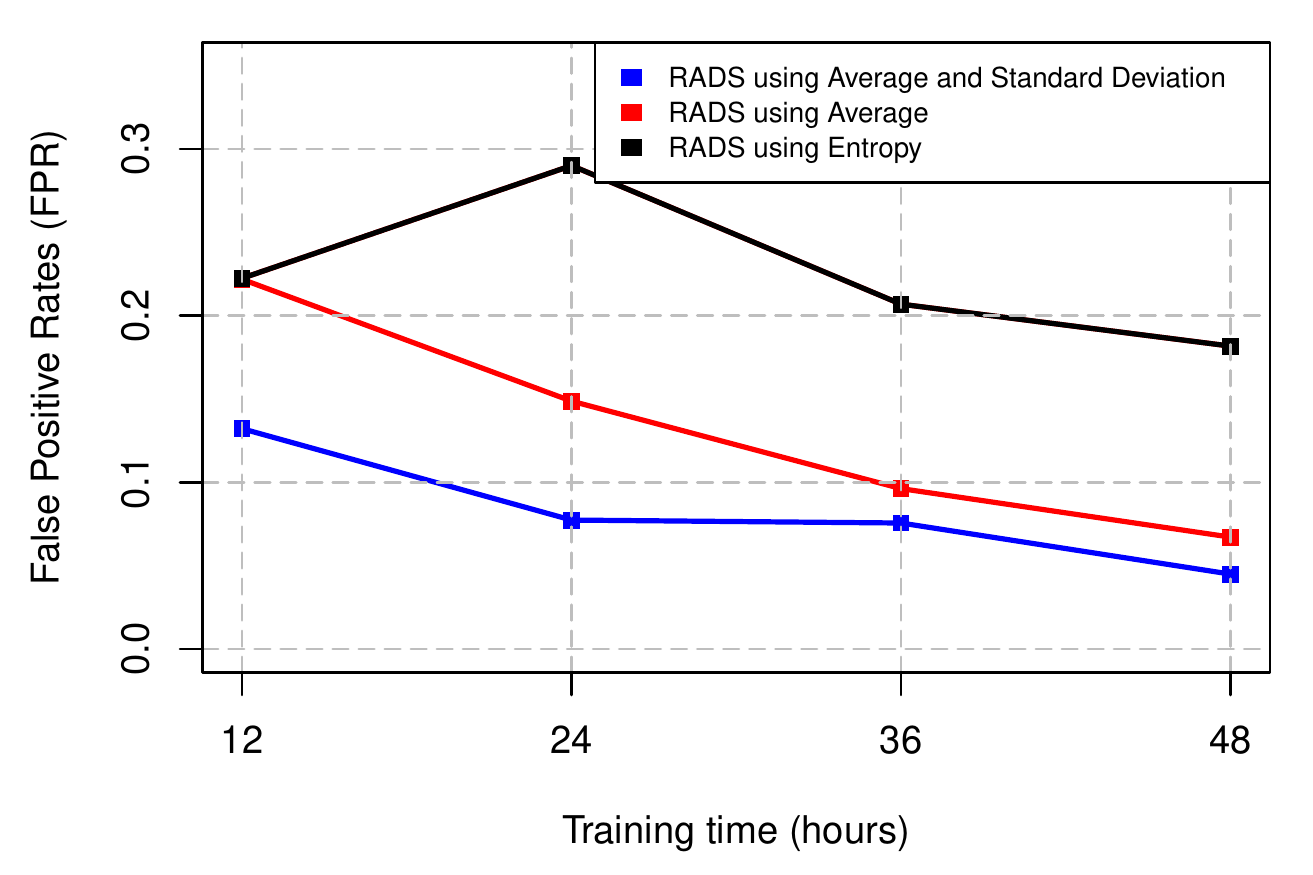, width = 0.9\columnwidth}}
   \caption{False Positive Rates (FPR) while running experiments on CPU utilisation (above) and network throughput (bottom)}
  \label{fig:offline_cpu_network}
\end{figure}

\textbf{Overall Performance of RADS:}
We evaluate the overall performance of RADS in terms of false positive rate. 
Figure~\ref{fig:offline_cpu_network} presents the results of the False Positive Rates (FPR, calculated using Equation~\ref{eq5}) of RADS while running the experiments on a CPU utilisation trace of 82 VMs and a network throughput trace of 212 VMs.
The experiments were performed with 24 hours of testing trace. 
We summarise the observations from the results as follows:
\begin{enumerate}[{(a)}]
\item On most occasions, when RADS uses its window-based time series analysis (combination of average and standard deviation), it achieves better performance (lower value of FPR means better performance) with increase in training time and at one stage (training time - from $36$ to $48$ hours) the performance starts becoming stable. These results emphasise further the requirement of the proposed training optimisation algorithm (Algorithm~\ref{raids_algorithm_training_optimiser}), which can decide the optimal training time.
\item The performance of RADS while using its window-based time series analysis is better than the performance of RADS while using average and entropy based analysis on most occasions.
\end {enumerate}


%

\section{Related Work}
\label{sec:related_work}
\noindent In recent years, researchers have proposed various anomaly detection systems for Cloud data centres. We classify them based on the machine learning algorithms which they implement.

\textit{(i) Supervised learning algorithms.}
Supervised learning algorithms rely on labelled training data to detect previously known anomalies. 
Li et al.~\cite{ml_based:2012} propose an Artificial Neural Network (ANN) based intrusion detection system for Cloud. The ANN algorithm learns the ``normal" and the ``anomalous" behaviour from a large dataset of VM network traffic. The learned ANN is capable of detecting Cloud security attacks with accurate results.
An anomaly detection system suitable for the hypervisor layer is proposed in~\cite{Pandeeswari2016}. The anomaly detection in this case is based on a mixture of Fuzzy C-Means clustering algorithm and Artificial Neural Network (FCM-ANN) which results in better accuracy and lower false positive rate than the classic ANN and Naive Bayes classifier for detecting various Cloud security attacks. 
The authors in~\cite{supervised_ml:2017} use Linear Regression (LR) and Random Forest (RF) algorithms to detect and categorise anomalies in a Cloud data centre. 
Gulenko et al.~\cite{ML_based_ids:2016} exploit various machine learning algorithms to detect anomalies in Cloud host machines. They use a combination of two types of data sets for evaluating the algorithms: ``normal" operation data and ``anomalous" data obtained via anomaly injection. They train the machine learning models offline and use them to detect the anomalies at runtime. 
The supervised learning algorithms used in Cloud anomaly detection as discussed above require training of the machine learning models with both ``normal" and ``anomalous" traces. 
These algorithms may fail to detect anomalies due to unknown attacks, traces of which are not recorded by the learning models or which have very different patterns than the learned ``anomalous" patterns.
To solve this problem researchers have proposed unsupervised learning algorithms which we discuss next. The unsupervised learning algorithms do not require labelled training data, i.e. they can build the learning models without the ``anomalous" traces.  

\textit{(ii) Unsupervised learning algorithms.}
The authors in~\cite{automated-detection:2016} propose a mechanism for automatic anomaly detection and root cause analysis for Cloud data centres. They use an unsupervised K-Means clustering algorithm to identify the ``abnormal" system behaviour. 
UBL proposed in~\cite{UBL:2012} uses an unsupervised Self Organising Map (SOM) algorithm to predict unknown anomalies. SOM is computationally less expensive than K-Nearest Neighbour~\cite{knearest:2005}. 
UBL predicts anomalies by identifying early deviations from ``normal" system behaviour. 
\cite{density-based:2016} proposes a Cloud anomaly detection technique based on the concept of data density introduced by~\cite{density-based_ref1:2011}, which implements non-parametric Cauchy function~\cite{density-based_ref2:2010}. 
This technique computes the density recursively and therefore, it is memory-less and unsupervised.  
The authors in~\cite{EbAT:2010}, \cite{entorpy_based_detection_2:2014} measure the entropy of the system metrics such as CPU, memory, network, IOPS, etc., in order to identify Cloud anomalies. The entropy values indicate the dispersal or concentration of the metric distributions and they form the time series data for anomaly analysis.
The approach proposed in~\cite{entorpy_based_detection_2:2014} identifies a Cloud security attack by observing whether the entropy variables obey normal distribution or not. They use Kolmogorov-Smirnov test (K-S test) to identify whether the entropy variables obey normal distribution. 
Recently, entropy has been used in various network anomaly detection tools~\cite{entorpy_based_detection_1:2005}, \cite{entorpy_based_detection_5:2014}, \cite{entorpy_based_detection_3:2017}, \cite{entorpy_based_detection_4:2017}. These tools firstly measure the entropy associated with the network traffic or network packet features (IP addresses and ports) and secondly they detect network attacks by observing the variation in the entropy values.
In our previous works~\cite{ladt:2015, ls-ladt:2016} we proposed a Lightweight Anomaly Detection Tool (LADT) which can detect anomalies on the hosting node level by using a correlation based algorithm. The algorithm utilises performance metrics on the hosting node level and the VM level to track disparities on the resource usage and detect host level attacks such as a Blue Pill attack~\cite{bluepill:2006}. However, this approach is not able to detect anomalies in the VM level which is the case for the current paper.

Although the unsupervised learning algorithms discussed above can detect Cloud anomalies due to unknown security attacks with high accuracy, they may generate false positives which arise mainly due to the workload spikes in a Cloud data centre.
The authors in~\cite{cloud-malware:2016} propose a novel approach for Cloud malware detection using one class Support Vector Machine (SVM) algorithm. One class SVM is an extension of the traditional two-class SVM, which was proposed by Sch\"{o}lkopf et al. in~\cite{one_class_svm:1999}. 
Similar to the OCC algorithm~\cite{OCC:2008} that is used in this paper, one class SVM takes the unlabelled training data and produces a binary class based on the distribution of the training data. The binary class is composed of a known class, which is the ``normal" VM behaviour and a novel class, which is the unknown class representing the ``anomalous" instances. 
The work in this paper is different from that in~\cite{cloud-malware:2016} as this work focuses more on increasing the accuracy while reducing the false positives arising due to genuine Cloud workload spikes; whereas, \cite{cloud-malware:2016} focuses on reducing false positives arising due to VM live-migration.

\section{Conclusion}
\label{sec:conclusions}
\noindent Cloud computing services have seen significant growth in recent years. Such growth has attracted various cybersecurity attacks on Cloud data centres. Reports from various security experts have raised concerns regarding the potential damage and growth of the cybersecurity attacks in the Cloud. 
Researchers have proposed a number of anomaly detection techniques to deal with such attacks. However, there exists some challenges, specifically due to the unknown behaviour of the attacks and the occurrence of genuine Cloud workload spikes.
In this paper, we discuss these challenges and investigate the issues with the existing Cloud anomaly detection approaches. Then, we propose a Real-time Anomaly Detection System (RADS) which uses One Class Classification (OCC) algorithm and a window-based time series analysis to address the challenges. 
We evaluate the performance of RADS by running lab-based and real-world experiments.
The lab-based experiments were performed in an OpenStack based Cloud data centre, which hosts two representative Cloud applications (Graph Analytics and Media Streaming) collected from the CloudSuite workload collection, whereas the real-world experiments were carried out on the real-world workload traces collected from a Cloud data centre named Bitbrains.
Evaluation results demonstrate that RADS can achieve 90-95\% accuracy (F1 score) with a low false positive rate of 0-3\% while detecting DDoS and cryptomining attacks in real-time. The results further reveal that RADS experiences fewer false positives while using the proposed window-based time series analysis than when using state-of-the-art average or entropy based analysis.
We also evaluate the efficiency of RADS in performing the training and the testing in real-time in our lab-based Cloud data centre while hosting varying numbers of VMs (2-10 VMs). 
The evaluation results suggest that RADS can be used as a lightweight tool in terms of consuming minimal hosting node CPU and processing time in a Cloud data centre.
However, to attain a more realistic evaluation of the efficiency of RADS, we need to perform the experiment with a significantly greater number of VMs.

\label{sec:conclusions}

\ifCLASSOPTIONcompsoc
  \section*{Acknowledgments}
\else
  \section*{Acknowledgment}
\fi
\noindent This work has received funding from the European Commission under the European Union's Seventh Framework Programme (grant agreement 610811 - CACTOS project) and the Horizon 2020 research and innovation programme (grant agreement 687628 - VINEYARD project), and the UK Engineering and Physical Sciences Research Council under grant agreements EP/L004232/1 - ENPOWER project and EP/M015742/1 - VINEYARD project.
\bibliographystyle{IEEEtran}  
\bibliography{references}

\begin{thebibliography}{10}
\providecommand{\url}[1]{#1}
\csname url@samestyle\endcsname
\providecommand{\newblock}{\relax}
\providecommand{\bibinfo}[2]{#2}
\providecommand{\BIBentrySTDinterwordspacing}{\spaceskip=0pt\relax}
\providecommand{\BIBentryALTinterwordstretchfactor}{4}
\providecommand{\BIBentryALTinterwordspacing}{\spaceskip=\fontdimen2\font plus
\BIBentryALTinterwordstretchfactor\fontdimen3\font minus
  \fontdimen4\font\relax}
\providecommand{\BIBforeignlanguage}[2]{{%
\expandafter\ifx\csname l@#1\endcsname\relax
\typeout{** WARNING: IEEEtran.bst: No hyphenation pattern has been}%
\typeout{** loaded for the language `#1'. Using the pattern for}%
\typeout{** the default language instead.}%
\else
\language=\csname l@#1\endcsname
\fi
#2}}
\providecommand{\BIBdecl}{\relax}
\BIBdecl

\bibitem{ml_based:2012}
Z.~Li, W.~Sun, and L.~Wang, ``A neural network based distributed intrusion
  detection system on cloud platform,'' in \emph{2012 IEEE 2nd International
  Conference on Cloud Computing and Intelligence Systems}, vol.~01, Oct 2012,
  pp. 75--79.

\bibitem{Pandeeswari2016}
\BIBentryALTinterwordspacing
N.~Pandeeswari and G.~Kumar, ``Anomaly detection system in cloud environment
  using fuzzy clustering based ann,'' \emph{Mobile Networks and Applications},
  vol.~21, no.~3, pp. 494--505, Jun 2016. [Online]. Available:
  \url{https://doi.org/10.1007/s11036-015-0644-x}
\BIBentrySTDinterwordspacing

\bibitem{ML_based_ids:2016}
A.~Gulenko, M.~Wallschläger, F.~Schmidt, O.~Kao, and F.~Liu, ``Evaluating
  machine learning algorithms for anomaly detection in clouds,'' in \emph{2016
  IEEE International Conference on Big Data (Big Data)}, Dec 2016, pp.
  2716--2721.

\bibitem{automated-detection:2016}
J.~Lin, Q.~Zhang, H.~Bannazadeh, and A.~Leon-Garcia, ``Automated anomaly
  detection and root cause analysis in virtualized cloud infrastructures,'' in
  \emph{NOMS 2016 - 2016 IEEE/IFIP Network Operations and Management
  Symposium}, April 2016, pp. 550--556.

\bibitem{UBL:2012}
\BIBentryALTinterwordspacing
D.~J. Dean, H.~Nguyen, and X.~Gu, ``Ubl: Unsupervised behavior learning for
  predicting performance anomalies in virtualized cloud systems,'' in
  \emph{Proceedings of the 9th International Conference on Autonomic
  Computing}, ser. ICAC '12.\hskip 1em plus 0.5em minus 0.4em\relax New York,
  NY, USA: ACM, 2012, pp. 191--200. [Online]. Available:
  \url{http://doi.acm.org/10.1145/2371536.2371572}
\BIBentrySTDinterwordspacing

\bibitem{cloud-malware:2016}
M.~R. Watson, N.~u.~h. Shirazi, A.~K. Marnerides, A.~Mauthe, and D.~Hutchison,
  ``Malware detection in cloud computing infrastructures,'' \emph{IEEE
  Transactions on Dependable and Secure Computing}, vol.~13, no.~2, pp.
  192--205, March 2016.

\bibitem{workloadCCGRID:2015}
S.~Shen, V.~v.~Beek, and A.~Iosup, ``Statistical characterization of
  business-critical workloads hosted in cloud datacenters,'' in \emph{2015 15th
  IEEE/ACM International Symposium on Cluster, Cloud and Grid Computing}, May
  2015, pp. 465--474.

\bibitem{EbAT:2010}
C.~Wang, V.~Talwar, K.~Schwan, and P.~Ranganathan, ``Online detection of
  utility cloud anomalies using metric distributions,'' in \emph{2010 IEEE
  Network Operations and Management Symposium - NOMS 2010}, April 2010, pp.
  96--103.

\bibitem{entorpy_based_detection_2:2014}
J.~Cao, B.~Yu, F.~Dong, X.~Zhu, and S.~Xu, ``Entropy-based denial of service
  attack detection in cloud data center,'' in \emph{2014 Second International
  Conference on Advanced Cloud and Big Data}, Nov 2014, pp. 201--207.

\bibitem{OCC:2008}
\BIBentryALTinterwordspacing
K.~Hempstalk, E.~Frank, and I.~H. Witten, ``One-class classification by
  combining density and class probability estimation,'' in \emph{Proceedings of
  the 2008 European Conference on Machine Learning and Knowledge Discovery in
  Databases - Part I}, ser. ECML PKDD '08.\hskip 1em plus 0.5em minus
  0.4em\relax Berlin, Heidelberg: Springer-Verlag, 2008, pp. 505--519.
  [Online]. Available: \url{http://dx.doi.org/10.1007/978-3-540-87479-9_51}
\BIBentrySTDinterwordspacing

\bibitem{ddso_charater_2017}
\BIBentryALTinterwordspacing
S.~Behal, K.~Kumar, and M.~Sachdeva, ``Characterizing ddos attacks and flash
  events: Review, research gaps and future directions,'' \emph{Computer Science
  Review}, vol.~25, pp. 101 -- 114, 2017. [Online]. Available:
  \url{http://www.sciencedirect.com/science/article/pii/S1574013717300941}
\BIBentrySTDinterwordspacing

\bibitem{entropy:2001}
\BIBentryALTinterwordspacing
C.~E. Shannon, ``A mathematical theory of communication,'' \emph{SIGMOBILE Mob.
  Comput. Commun. Rev.}, vol.~5, no.~1, pp. 3--55, Jan. 2001. [Online].
  Available: \url{http://doi.acm.org/10.1145/584091.584093}
\BIBentrySTDinterwordspacing

\bibitem{supervised_ml:2017}
T.~Salman, D.~Bhamare, A.~Erbad, R.~Jain, and M.~Samaka, ``Machine learning for
  anomaly detection and categorization in multi-cloud environments,'' in
  \emph{2017 IEEE 4th International Conference on Cyber Security and Cloud
  Computing (CSCloud)}, June 2017, pp. 97--103.

\bibitem{knearest:2005}
P.-N. Tan, M.~Steinbach, and V.~Kumar, \emph{Introduction to Data Mining,
  (First Edition)}.\hskip 1em plus 0.5em minus 0.4em\relax Boston, MA, USA:
  Addison-Wesley Longman Publishing Co., Inc., 2005.

\bibitem{density-based:2016}
S.~N. Shirazi, S.~Simpson, A.~Gouglidis, A.~Mauthe, and D.~Hutchison, ``Anomaly
  detection in the cloud using data density,'' in \emph{2016 IEEE 9th
  International Conference on Cloud Computing (CLOUD)}, June 2016, pp.
  616--623.

\bibitem{density-based_ref1:2011}
P.~Angelov and R.~Yager, ``Simplified fuzzy rule-based systems using
  non-parametric antecedents and relative data density,'' in \emph{2011 IEEE
  Workshop on Evolving and Adaptive Intelligent Systems (EAIS)}, April 2011,
  pp. 62--69.

\bibitem{density-based_ref2:2010}
N.~Luo and F.~Qian, ``Estimation of distribution algorithm sampling under
  gaussian and cauchy distribution in continuous domain,'' in \emph{IEEE ICCA
  2010}, June 2010, pp. 1716--1720.

\bibitem{entorpy_based_detection_1:2005}
\BIBentryALTinterwordspacing
A.~Lakhina, M.~Crovella, and C.~Diot, ``Mining anomalies using traffic feature
  distributions,'' \emph{SIGCOMM Comput. Commun. Rev.}, vol.~35, no.~4, pp.
  217--228, Aug. 2005. [Online]. Available:
  \url{http://doi.acm.org/10.1145/1090191.1080118}
\BIBentrySTDinterwordspacing

\bibitem{entorpy_based_detection_5:2014}
L.~Zhao and F.~Wang, ``An efficient entropy-based network anomaly detection
  method using mib,'' in \emph{2014 IEEE International Conference on Progress
  in Informatics and Computing}, May 2014, pp. 428--432.

\bibitem{entorpy_based_detection_3:2017}
\BIBentryALTinterwordspacing
S.~Behal and K.~Kumar, ``Detection of ddos attacks and flash events using novel
  information theory metrics,'' \emph{Comput. Netw.}, vol. 116, no.~C, pp.
  96--110, Apr. 2017. [Online]. Available:
  \url{https://doi.org/10.1016/j.comnet.2017.02.015}
\BIBentrySTDinterwordspacing

\bibitem{entorpy_based_detection_4:2017}
C.~Callegari, S.~Giordano, and M.~Pagano, ``Entropy-based network anomaly
  detection,'' in \emph{2017 International Conference on Computing, Networking
  and Communications (ICNC)}, Jan 2017, pp. 334--340.

\bibitem{ladt:2015}
S.~Barbhuiya, Z.~Papazachos, P.~Kilpatrick, and D.~S. Nikolopoulos, ``A
  lightweight tool for anomaly detection in cloud data centres,'' in
  \emph{Proceedings of the 5th International Conference on Cloud Computing and
  Services Science}, 2015, pp. 343--351.

\bibitem{ls-ladt:2016}
\BIBentryALTinterwordspacing
------, \emph{LS-ADT: Lightweight and Scalable Anomaly Detection for Cloud
  Datacentres}.\hskip 1em plus 0.5em minus 0.4em\relax Cham: Springer
  International Publishing, 2016, pp. 135--152. [Online]. Available:
  \url{https://doi.org/10.1007/978-3-319-29582-4_8}
\BIBentrySTDinterwordspacing

\bibitem{bluepill:2006}
J.~Rutkowska, ``Subverting vistatm kernel for fun and profit,'' in \emph{Black
  Hat Conference}, Sept 2006.

\bibitem{one_class_svm:1999}
\BIBentryALTinterwordspacing
B.~Sch\"{o}lkopf, R.~Williamson, A.~Smola, J.~Shawe-Taylor, and J.~Platt,
  ``Support vector method for novelty detection,'' in \emph{Proceedings of the
  12th International Conference on Neural Information Processing Systems}, ser.
  NIPS'99.\hskip 1em plus 0.5em minus 0.4em\relax Cambridge, MA, USA: MIT
  Press, 1999, pp. 582--588. [Online]. Available:
  \url{http://dl.acm.org/citation.cfm?id=3009657.3009740}
\BIBentrySTDinterwordspacing

\end{thebibliography}

\begin{IEEEbiography}
[{\includegraphics[width=1in,height=1.25in,clip,keepaspectratio]{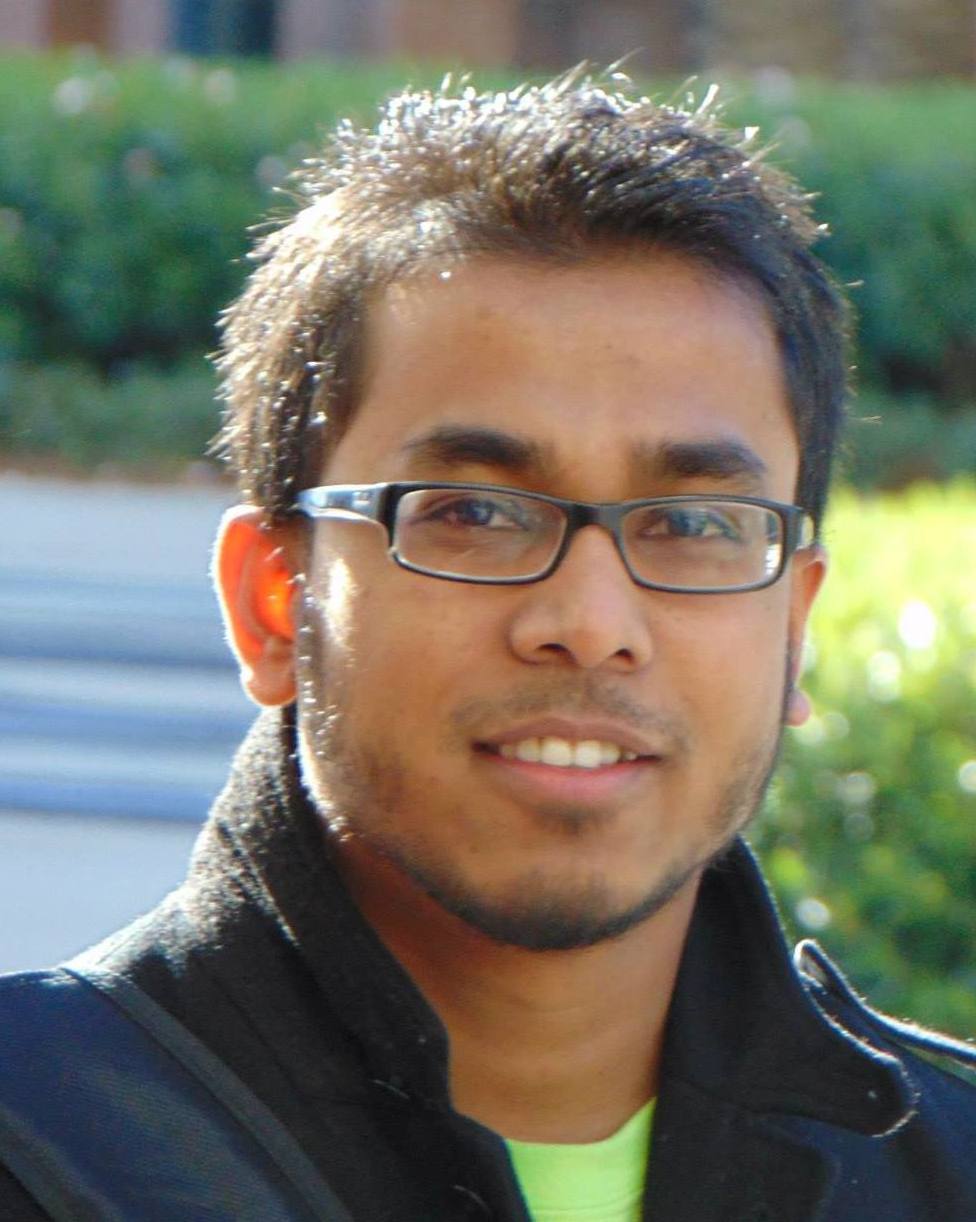}}]{Sakil Barbhuiya} completed his PhD in early 2018 in Computer Science from Queen’s University Belfast. His PhD research delivered accurate and efficient anomaly detection techniques for Cloud and mobile devices to protect them against cybersecurity attacks. During his PhD Sakil was part of couple of EU-funded projects, namely CACTOS and VINEYARD.
Currently, Sakil is a postdoctoral research fellow in the Centre for Data Science and Scalable Computing in the School of Electronics, Electrical  Engineering and Computer Science at Queen's University Belfast.
Overall, Sakil's research interests include Cloud computing, data analytics, machine learning, and anomaly detection. 
\end{IEEEbiography}

\begin{IEEEbiography}
[{\includegraphics[width=1in,height=1.25in,clip,keepaspectratio]{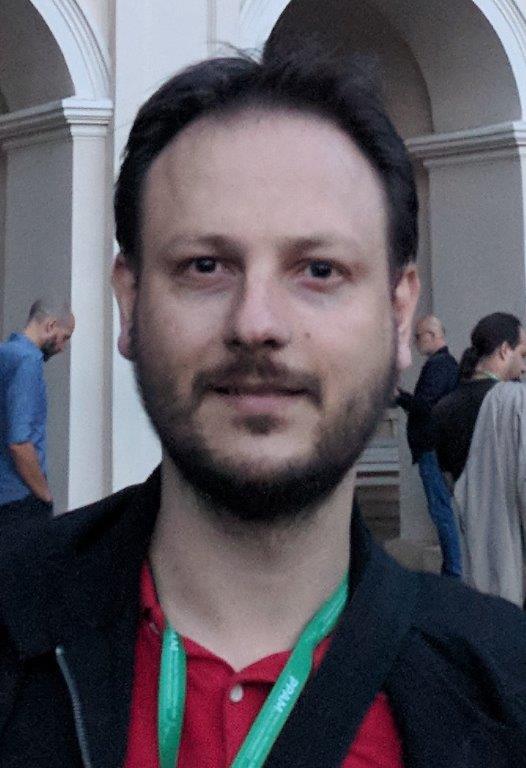}}]{Zafeirios Papazachos} is a postdoctoral research fellow in the Centre for 
Data Science and Scalable Computing in the School of Electronics, Electrical  
Engineering and Computer Science at Queen's University Belfast. His research interests include scalable methods for
big data monitoring, collection and analysis; performance evaluation of Cloud and distributed systems, scheduling, modelling and simulation. His current 
research area also includes efficient integration and utilisation of accelerators in large scale distributed systems and Cloud environments.
\end{IEEEbiography}

\begin{IEEEbiography}
[{\includegraphics[width=1in,height=1.25in,clip,keepaspectratio]{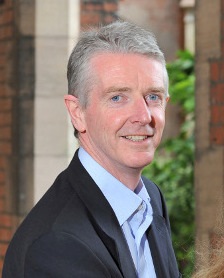}}]{Peter Kilpatrick} is a member of the Centre for Data Science and Scalable Computing in the School of Electronics, Electrical  Engineering and Computer Science at Queen's University Belfast. His research interests include programming models for parallel/distributed systems and management of non-functional concerns in such systems. He is the author of over 120 papers in peer-reviewed  international conferences and journals. He has participated in a  number of EU-funded projects including Cactos (2013-2016), ParaPhrase (2011-2015) and CoreGRID (2004-2008) and currently is a member of the EU COST Action cHiPSet (2015-2019). He is UK Director for Euromicro and is a member of the Register of Expert Peer Reviewers for Italian Scientific Evaluation (REPRISE).
\end{IEEEbiography}

\begin{IEEEbiography}
[{\includegraphics[width=1in,height=1.25in,clip,keepaspectratio]{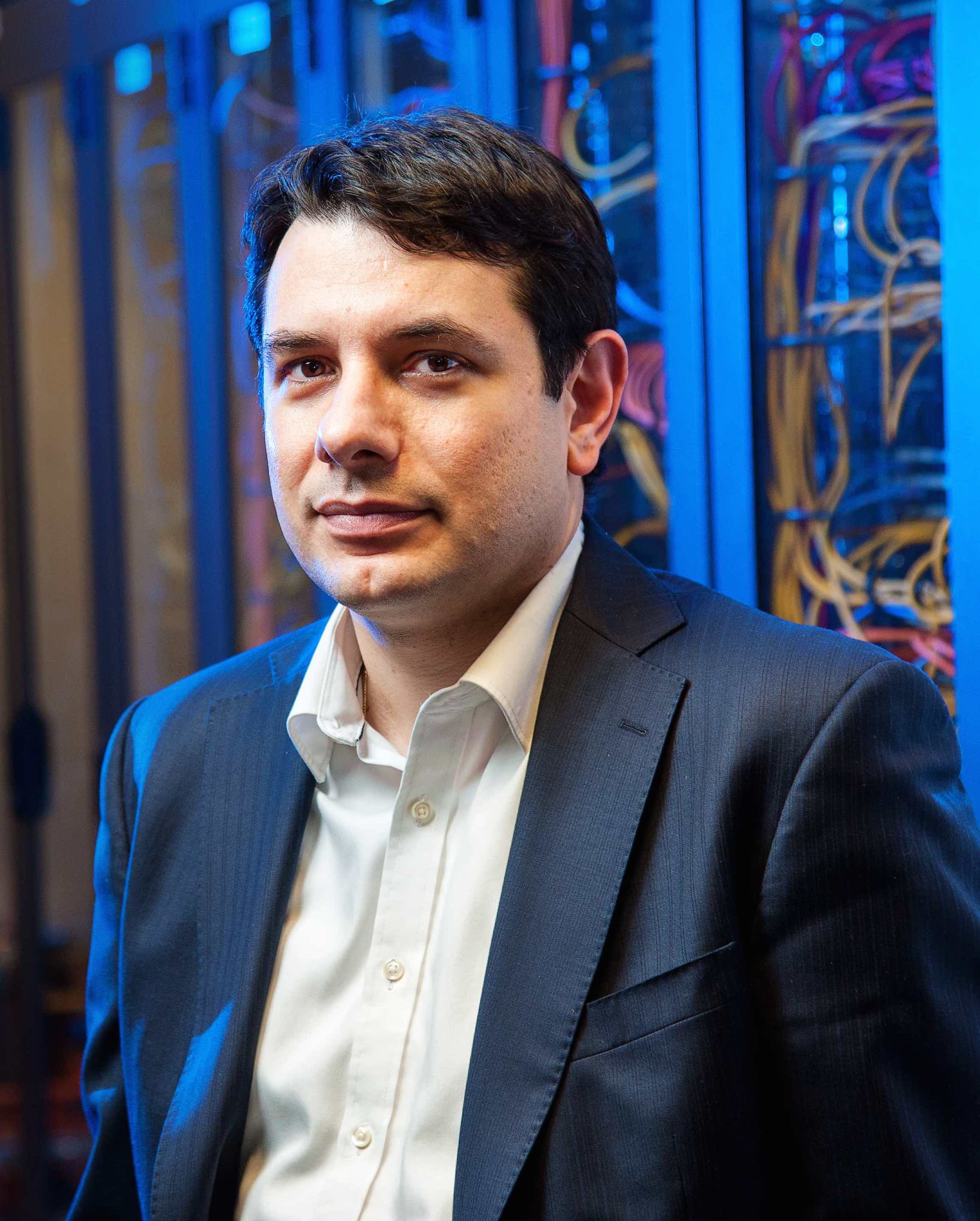}}]{Dimitrios S. Nikolopoulos} FBCS FIET is Professor in the School of Electronics, Electrical
Engineering and Computer Science, and Director of the Global Research Institute
on Electronics, Communication and Information Technologies at Queen's
University Belfast. He holds the Chair in High Performance and Distributed Computing and is also
a Royal Society Wolfson Research Fellow. His
research explores scalable computing systems for data-driven applications and new computing
paradigms at the limits of performance, power and reliability. Dimitrios holds BEng (96), MEng (97) and PhD degrees (00)
in Computer Engineering from the University of Patras. 
\end{IEEEbiography}

\end{document}